\documentclass[12pt,letterpaper]{article} 

\usepackage[includeheadfoot,
            marginratio={1:1,2:3}, 
            width=412pt, 
            height=688pt,]{geometry}

\usepackage{amsmath}
\usepackage{amsfonts}
\usepackage{amssymb}
\usepackage{stmaryrd}
\usepackage{dsfont}
\usepackage{graphicx}
\usepackage{amscd}
\usepackage{extarrows}
\usepackage{hyperref}
\usepackage[all]{hypcap}

\newcommand{\nc}{\newcommand}
\nc{\lb}{\llbracket}
\nc{\rb}{\rrbracket}
\nc{\gl}{\llbracket}
\nc{\gr}{\rrbracket}

\newcommand{\eq}[1]{\begin{equation}
                     \begin{split} #1 \end{split}
                     \end{equation}}

\newcommand{\ov}{\overline}

\allowdisplaybreaks[2]
\numberwithin{equation}{section}

\newcommand{\fa}[1]{#1}

\newcommand{\td}[1]{$\rho$-#1}
\newcommand{\mdet}[1]{\left|#1\right|}   

\begin{document}

\vspace*{-1.5cm}

\begin{flushright}
  {\small
  MPP-2013-99\\
  DFPD-2013-TH-06
  }
\end{flushright}

\vspace{1.5cm}


\begin{center}
{\LARGE
The Intriguing Structure of Non-geometric \\[0.25cm]
Frames in String Theory
}
\end{center}


\vspace{0.4cm}

\begin{center}
  Ralph Blumenhagen$^{1}$, Andreas Deser$^{1}$, Erik Plauschinn$^{2,3}$, \\[0.1cm]
Felix Rennecke$^{1}$ and Christian Schmid$^{1}$
\end{center}


\vspace{0.4cm}

\begin{center} 
\emph{$^{1}$ Max-Planck-Institut f\"ur Physik (Werner-Heisenberg-Institut), \\ 
   F\"ohringer Ring 6,  80805 M\"unchen, Germany } \\[0.1cm] 
\vspace{0.25cm}
\emph{$^{2}$ Dipartimento di Fisica e Astronomia ``Galileo Galilei'' \\
Universit\`a  di Padova, Via Marzolo 8, 35131 Padova, Italy}  \\[0.1cm] 
\vspace{0.25cm}
\emph{$^{3}$ INFN, Sezione di Padova \\
Via Marzolo 8, 35131 Padova, Italy}  \\
\end{center} 

\vspace{0.4cm}


\begin{abstract}
Non-geometric frames in string theory are related 
to the geometric ones by certain local $O(D,D)$ transformations,
the so-called $\beta$-transforms. 
For each such  transformation, 
we show that there exists both a natural field redefinition
of the metric and the Kalb-Ramond two-form as well as
an associated Lie algebroid.
We furthermore prove that the all-order low-energy effective action of the 
superstring, written in terms of the redefined fields,
can be expressed through differential-geometric objects of the corresponding Lie algebroid.
Thus, the latter  provides a natural framework for  
effective superstring actions in non-geometric frames.
Relations of this new formalism to double field theory
and to the description of non-geometric backgrounds such as T-folds are
discussed as well.
\end{abstract}


\clearpage

\tableofcontents


\section{Introduction}
\label{sec:intro}

One of the celebrated  features of string theory is that 
after quantizing the closed string, one generically finds
a massless mode in the spectrum, which has all the properties  of
a graviton. Another important aspect is that the graviton
is accompanied by two additional massless excitations,
namely the  Kalb-Ramond field  and the dilaton.
The leading-order dynamics of these fields is governed
by an effective action containing the Einstein-Hilbert term
for gravity and the kinetic terms of 
the Kalb-Ramond field and the dilaton. 
This action in the so-called geometric frame
has two types of local symmetries, namely 
it is invariant  under diffeomorphisms of the space-time
coordinates and under gauge transformations of the Kalb-Ramond field.
String theory furthermore provides higher-order $\alpha'$-corrections
which involve e.g.\ higher
powers of the Riemann tensor.

String theory transcends the usual notions of field theory
by the existence of new transformations where string
momentum and winding modes are exchanged. 
These so-called T-dualities are crucial and have been
a valuable guide  for the detection
of new structures in string theory, such as mirror symmetry
or D-branes. Moreover, this T-duality, via the Buscher
rules, acts non-trivially on the metric, the Kalb-Ramond form
and the dilaton. In particular the metric and the Kalb-Ramond
field become closely intertwined. 
For a compactification on a $D$-dimensional torus, the $D^2$-dimensional
moduli space becomes $O(D,D;\mathbb{R})/O(D)\times O(D)$  which in string theory is further
divided by the T-duality group $O(D,D;\mathbb{Z})$.

In view of this, it is a natural question  whether one can implement these
$O(D,D)$ transformations, whose origin lies in the decoupling
of left- and right-movers on the string world-sheet,
directly in the space-time effective action of string theory. 
Indeed, following  some earlier work \cite{Siegel:1993th,Siegel:1993xq}, two frameworks 
were developed where the $O(D,D)$ transformations \footnote{If not otherwise
  specified, the short-hand notation $O(D,D)$ stands for \emph{local} $O(D,D)$
  transformations, i.e.\ those which non-trivially depend on the coordinates.}
play
a crucial role, namely generalized 
geometry \cite{Hitchin:2004ut,Gualtieri:2003dx,Grana:2008yw,Coimbra:2011nw}
and  double field theory (DFT) \cite{Hull:2009mi,Hull:2009zb,Hohm:2010jy,Hohm:2010pp,Aldazabal:2011nj}.
In the first approach, the concept of Riemannian geometry is extended 
from the tangent bundle $TM$ to the generalized tangent bundle
$TM\oplus T^*M$, whereas in the second
the dimension of the space is doubled by including winding coordinates subject
to certain constraints.
For the latter construction, this admits a manifest global 
 $O(D,D)$  invariance of the
action, so in particular, 
the action is manifestly invariant under T-duality transformations.
The fundamental object in both approaches is a generalized metric
which combines the usual metric and Kalb-Ramond field.
The two local symmetries, diffeomorphisms and $B$-field gauge transformations,
sit inside a subgroup of $O(D,D)$.
Their complement in $O(D,D)$ 
contains so-called (local) $\beta$-transforms, which 
lead out of the usual geometric
frame  of string theory. Therefore, applying a local $\beta$-transform to the 
geometric frame leads to what we call a {\it non-geometric frame}.

The existence of {\it non-geometric backgrounds} can be seen by analyzing the 
action of T-duality on the simple background of a flat three-dimensional
torus  with a constant $H$-flux \cite{Shelton:2005cf}. 
Applying successive T-dualities,
this $H$-flux is first mapped to a geometric flux \cite{Kachru:2002sk} and
by a second T-duality to the non-geometric $Q$-flux \cite{Hellerman:2002ax,Dabholkar:2002sy,Hull:2004in}.
The latter background can be understood as a T-fold \cite{Hull:2006va}, where the
transition functions between two charts involve
T-duality transformations. A third T-duality is beyond the
scope of the Buscher rules, and both non-com\-mu\-ta\-tive geometry
\cite{Bouwknegt:2004ap,Mylonas:2012pg,Chatzistavrakidis:2012qj} 
and conformal field theory
\cite{Blumenhagen:2010hj,Lust:2010iy,Blumenhagen:2011ph,Condeescu:2012sp,Andriot:2012vb} hint
towards a non-associative structure. The effect of T-duality on brane
solutions
has been analyzed recently in \cite{Hassler:2013wsa}.

Since in DFT a global $O(D,D)$ symmetry is manifest,
the first-order effective action in at least a subset of these non-geometric
frames is also described by it. What has been puzzling
is that the DFT  action cannot be straightforwardly interpreted
as the Einstein-Hilbert action of some  $O(D,D)$ 
covariant differential geometry \cite{Hohm:2012mf,Berman:2013uda}. 
The problem is
that the notions of torsion and curvature have to be changed
to make them tensors so that they do not satisfy
some of the usual properties of Riemannian geometry --
the Levi-Civita connection is not unique and the
curvature has more symmetries compared to the usual case.
That is not a major problem in itself, but higher-order $\alpha'$-corrections
involve the full Riemann tensor, so it is not clear how to describe these. 
The analogous situation  has also been encountered
in attempts to generalize DFT to M-theory  by making  the U-duality groups 
manifest  (see e.g.\ \cite{Berman:2010is,Coimbra:2011ky,Aldazabal:2013mya}).

In this paper we follow a slightly  less ambitious approach
which is motivated by the recent studies of effective
actions in non-geometric frames.
In \cite{Andriot:2011uh,Andriot:2012wx,Andriot:2012an} the 
geometric action was redefined using a non-geometric frame.
This
gave an action containing the metric and a bi-vector
field $\beta$ as the dynamical fields and involved
a new type of Ricci scalar. In \cite{Blumenhagen:2012nk,Blumenhagen:2012nt} 
the starting point was 
the abstract structure of a Lie algebroid and,
for a special case, a differential geometry was developed
whose Einstein-Hilbert term could be related 
to the Einstein-Hilbert term in the geometric frame
via a field redefinition. At that stage these two
approaches might look a bit ad hoc.

We clarify the conceptual status
of these two actions and show that they fit into a larger picture
in which mathematically the differential geometry 
of Lie algebroids plays an important role.
The starting point is the geometric frame. Then,  applying
a general local $O(D,D)$ transformation, from its action
on the generalized metric we can read off a field
redefinition for the metric and $B$-field.
For the geometric subgroup of diffeomorphisms and gauge transformations
this reduces to the familiar  form, however $\beta$-transformations
give a non-trivial redefinition.
With the field redefinition at hand,  one can express the
action in terms of these new field variables.
We show that for each non-geometric local $O(D,D)$ transformation
this action is based on nothing else
than the differential geometry of a corresponding Lie algebroid,
whose defining data can also be directly read off from the
$O(D,D)$ matrix.

Thus, this allows us to describe the low-energy effective action
of string theory in every non-geometric frame in terms
of a (generalized) differential geometry where, opposed to DFT,
the definitions  of torsion and curvature still keep
the familiar forms. Therefore, there also exists a 
Riemann tensor and it is clear how higher-order    
$\alpha'$-corrections are described in these non-geometric frames.
To emphasize it again, we are not, as in DFT, covariantizing part of the entire
$O(D,D)$ symmetry, but provide  a uniform description of
the string actions in any non-geometric frame in terms of a new differential
geometry. In each such frame, the action only enjoys the
usual diffeomorphism and gauge symmetries. 
Working still in the framework of generalized geometry, in contrast  to DFT,
we do not have the local symmetries related to the winding-coordinate 
dependence of the usual and winding diffeomorphisms. 
As we will see, as a consequence, the description of global 
non-geometric backgrounds,
like the constant $Q$-flux example, is not possible within a single 
frame.\footnote{Note that in DFT, a non-geometric background can be characterized by
the appearance of winding coordinates either directly in the
dependence of the DFT metric (as in the toroidal constant $R$-flux example)  
or  in the transition functions between two patches
(as in the toroidal constant $Q$-flux example).}

This paper is organized as follows: In section 2 we recall some
basics notions of generalized geometry and show that every
$O(D,D)$ transformation naturally induces a corresponding field redefinition.
For $\beta$-transformations, this goes beyond  the realm of differential geometry. 
Two examples are presented, which were
previously discussed in the literature.
We point out that the mathematical framework, capturing the structure
of the geometry in the redefined variables, is based on so-called
Lie algebroids.
In section 3, after an introduction to Lie algebroids we outline
the corresponding differential geometry which 
by construction is covariant under diffeomorphisms.
Then we discuss how one can define also a Lie algebroid from
an $O(D,D)$ transformation. In section 4, we generally prove that
the differential geometry in the redefined variables is nothing
else than the differential geometry of the corresponding Lie algebroid.
The final NS-NS action in the redefined variables is presented
and shown to be invariant under diffeomorphisms and
the analog of $B$-field gauge transformations in the new variables.
In section 5 we discuss further aspects of this formalism, namely
we clarify the relation to double field theory, the extension
to superstring effective actions to higher-order
$\alpha'$-corrections and provide the tree-level equations
of motions in each non-geometric frame. Finally, we elaborate
on  the relation and distinction between what we have called
non-geometric frames, which is a choice of variables, and
the description of global non-geometric string backgrounds.
The upshot is
that, in a non-geometric frame, in each patch a non-geometric
background might take a very simple form. However,
the transition functions are still given by transformations, i.e.\
$\beta$-transforms, which are not a symmetry of the action in each patch.


\section{Generalized geometry}

In this section, we show that for every local $O(D,D)$ transformation a corresponding 
field redefinition can be deduced. In order to do so, we start by recalling  some basics on 
generalized geometry.


\subsection{$O(D,D)$ transformations and the generalized metric}
\label{sec_odd}

Let us  briefly introduce $O(D,D)$ transformations as well as the concept of 
a generalized metric. For more details, we refer the reader to 
\cite{Grana:2008yw}.


\subsubsection*{Basics on generalized geometry}

We consider a $D$-dimensional manifold $M$ together 
with the so-called generalized tangent bundle  $E={TM}\oplus {T^*M}$. 
The elements in $E$ will be denoted by the formal sum $(X+\xi)\in\Gamma(E)$, 
where $X\in \Gamma(TM)$ is a vector field and $\xi\in \Gamma(T^*M)$ is a one-form.
The natural bilinear form on the bundle $E$ is 
\eq{
  \label{bilinearform}
  \langle X+\xi,Y+\zeta \rangle = \xi(Y)+ \zeta(X) \,,
}
where the action of say $\xi=\xi_{\alpha} e^{\alpha}$ on $Y=Y^a e_a$ is given by $\xi(Y)=\xi_a Y^a$.
The bilinear form \eqref{bilinearform} can also be described in terms of a $2D\times 2D$ matrix\,\footnote{Explicitly, this means that \eqref{bilinearform} can be written as 
$    \langle X+\xi,Y+\zeta \rangle  = \scriptsize\renewcommand{\arraystretch}{0.9}\arraycolsep1pt 
\left( \begin{array}{c} X^a \\ \xi_{\alpha} \end{array} \right)^t \!
\left(\begin{array}{cc} 0& \delta_a{}^{\beta}\\ \delta^{\alpha}{}_{b}&0\end{array}\right)\left(\begin{array}{c}Y^b \\ \zeta_{\beta} \end{array}\right)$.
}
\eq{
  \label{bilinearform_02}
   \eta= \begin{pmatrix} 0& \mathds 1\\ \mathds 1&0\end{pmatrix}.
}
The transformations $\mathcal M$ which leave \eqref{bilinearform_02} invariant, that is
\eq{
\label{odd}
	\mathcal M^t\, \eta\, \mathcal M = \eta \,,
}
constitute the group $O(D,D)$. A general matrix $\mathcal M\in O(D,D)$ 
can be decomposed into four $D\times D$ matrices as follows
\eq{
  \label{odd_trafo_02}
    \mathcal M=\begin{pmatrix} a&b\\c&d\end{pmatrix} \,,
}
and equation \eqref{odd} then yields three independent constraints on the submatrices,
namely
\eq{\label{odds}
	a^t c +c^t a &=0 \,,\\	b^t d +d^t b &=0 \,,\\
  b^t c +d^t a &=\mathds 1 \,.
}
Note that in our conventions, the $O(D,D)$ matrix \eqref{odd_trafo_02} acts on a tuple 
$(X^a,\xi_{\alpha})^t$, 
with $X=X^a e_a$ a vector field and $\xi=\xi_{\alpha} e^{\alpha}$ a one-form.
Therefore, the index structure of the submatrices in \eqref{odd_trafo_02} is
\eq{	
  \label{index_structure}
  a^a{}_{b}\,, \hspace{20pt}
  b^{a\beta}\,, \hspace{20pt}
  c_{\alpha b}\,, \hspace{20pt}
  d_{\alpha}{}^{\beta}\,.
}


\subsubsection*{The generalized metric}

Let us now combine the metric $G_{ab}$ of the manifold $M$ and the antisymmetric Kalb-Ramond field $B_{ab}$ into the so-called generalized metric 
\eq{
\label{genmetric}
    {\cal H}=
      \left(\begin{matrix}  G -BG^{-1}B& BG^{-1} \\
                                 -G^{-1}B  & G^{-1} \end{matrix}\right)   .
}
Note that $\mathcal H$ satisfies $( \eta \mathcal H )^2 = \mathds 1$,
and that elements of the group $O(D,D)$ act on the generalized 
metric by conjugation
\eq{ 
  \label{odd_trafo_01}
         \widehat{\cal H}=\mathcal M^t\, {\cal H}\, \mathcal M \;, \hspace{60pt}
         \mathcal M \in O(D,D) \,.
}
Since in general the  metric $\mathcal H$ depends non-trivially on the
coordinates $x \in M$ through $G$ and $B$, we  allow for an $x$-dependence 
in the transformation matrix,
i.e.\ we consider local $O(D,D)$ transformations  ${\cal M}(x)$. However, to keep our formulas 
readable, we mostly omit the explicit coordinate dependence in the following.

Since $G$ is symmetric and $B$ is antisymmetric, a priori $\mathcal H$ contains $D^2$ free parameters. 
But because $O(D,D)$  has $2 D^2- D$ free parameters, it is suggestive that there exists a 
subgroup 
of $O(D,D)$ 
 which leaves ${\cal H}$ invariant. These automorphisms are represented by the matrices
\eq{\label{Mauto}
	\mathcal{M}_\mathrm{auto}^{(1)} &=
	\begin{pmatrix} \mathcal O_1 & 0 \\ B\hspace{1pt} \mathcal O_1-( \mathcal O_1^t)^{-1}B \;
	& (\mathcal O_1^t)^{-1}
	\end{pmatrix} , \\[1.5mm]
	\mathcal{M}_\mathrm{auto}^{(2)} &=
	\begin{pmatrix} -G^{-1}(\mathcal O_2^t)^{-1}B & G^{-1}(\mathcal O_2^t)^{-1} 
	\\ G\hspace{1pt} \mathcal O_2-BG^{-1}(\mathcal O_2^t)^{-1}B\; & BG^{-1}(\mathcal O_2^t)^{-1}
	\end{pmatrix},
}
where $\mathcal O_{1,2} \in O_G(D)$.\footnote{We consider the local orthogonal group with respect to the metric $G$,
that is those matrices $\mathcal O$ which satisfy $\mathcal O^t G \mathcal O = G$. The metric is  positive definite as we are considering a Euclidean manifold $M$.}
It can be checked explicitly that transformations of the form \eqref{Mauto} preserve the generalized metric  \eqref{genmetric}.


\subsubsection*{$O(D,D)$ transformations}

Let us now turn to other subgroups of $O(D,D)$, which will become important in our subsequent discussion.
\begin{itemize}

\item The geometric subgroup $G_{\rm geom}\subset O(D,D)$ consists
of the group of diffeomorphisms $G_{\rm diffeo}\subset G_{\rm geom}$
characterized by
\eq{
\label{diffeos}
           \mathcal  M_{\rm diffeo}=\left(\begin{matrix} \mathsf  A & 0 \\
                                 0  & (\mathsf A^t)^{-1} \end{matrix}\right) ,
}
with $\mathsf A$ an invertible  $D\times D$ matrix. The matrices \eqref{diffeos} 
give rise to diffeomorphism transformations of the metric and $B$-field, which can be seen from
\eq{
  \mathcal H \bigl(\, \mathsf A^tG\hspace{1pt}\mathsf A, A^tB\hspace{1pt}\mathsf A \,\bigr) = \mathcal M_{\rm diffeo}^t\, \mathcal H(G,B) 
  \, \mathcal M_{\rm diffeo} \,.
}

\item The group of so-called {\em $B$-transforms} $G_{\mathsf B}\subset O(D,D)$ is given by matrices
\eq{
\label{Btrans}
     \mathcal  M_{\mathsf B}=\left(\begin{matrix} \mathds 1 & 0 \\
                                 -\mathsf B  & \mathds 1 \end{matrix}\right),
}
where $\mathsf B$ is an antisymmetric $D\times D$ matrix. For $\mathsf B = d \Lambda$, these 
$B$-transforms describe gauge transformations
of the Kalb-Ramond field. Indeed, one can check that
\eq{
  \mathcal H (G, B + d\Lambda) = \mathcal M_{d \Lambda}^t\, \mathcal H(G,B) 
  \, \mathcal M_{d\Lambda} \,.
}
The latter transformations therefore belong to the geometric subgroup $G_{\rm geom}$,
in particular, $G_{d\Lambda}$ is a normal subgroup of $G_{\rm geom}$, i.e.\
$G_{\rm geom}=G_{d\Lambda}\rtimes G_{\rm diffeo}$.

\item Finally, the so-called {\em $\beta$-transforms} $G_{\beta}$ are contained in
the complement $O(D,D)/G_{\rm geom}$ and take the form
\eq{\label{betatrans}
     \mathcal M^{\beta}=\left(\begin{matrix}  \mathds 1 & -\beta \\
                                 0  & \mathds 1 \end{matrix}\right),
}
whose action on $\mathcal H$
is not just given by diffeomorphisms or gauge transformations, but 
goes beyond  the geometric frame. 
Hence, the resulting new frame is  called a {\em non-geometric frame}.

\end{itemize}
In table \ref{tab:transf}, we have summarized the three types of transformations discussed in this paragraph.

\begin{table}[t]
\begin{center}
\begin{tabular}{|l@{\hspace{30pt}}l@{\hspace{2.5pt}}l@{\hspace{30pt}}r|}
\hline
&&&\\[-2.2mm]
diffeomorphisms & 
$            \mathcal  M_{\rm diffeo}$&$=\left(\begin{matrix} \mathsf  A & 0 \\
                                 0  & (\mathsf A^t)^{-1} \end{matrix}\right) $ &
$G_{\rm diffeo} \subset G_{\rm geom} \subset  O(D,D)$
\\[5mm]
$B$-transforms &
$     \mathcal  M_{\mathsf B}$&$=\left(\begin{matrix} \mathds 1 & 0 \\
                                 -\mathsf B  & \mathds 1 \end{matrix}\right) $ &
$G_{\mathsf B}  \subset  O(D,D)$ 
\\[5mm]
$\beta$-transforms &
$    \mathcal   M^{\beta}$&$=\left(\begin{matrix}  \mathds 1 & -\beta \\
                                 0  & \mathds 1 \end{matrix}\right)
$ &
$G_{\beta} \subset O(D,D)/G_{\rm geom}$
\\[5mm]
\hline
\end{tabular}                               
\caption{\label{tab:transf}Summary of  $O(D,D)$ transformations discussed in the main text.}
\end{center}  
\end{table}


\subsection{$O(D,D)$-induced field redefinition}
\label{sec_oddredef}

As we have illustrated, the generalized metric \eqref{genmetric} encodes $G$ and $B$
in a way  that  is suitable for implementing the $O(D,D)$ structure. 
However, a general $O(D,D)$ transformation mixes the entries of $\mathcal{H}(G,B)$ in a 
complicated manner. 
If we want to cast the transformed metric $\widehat{\mathcal{H}}(G,B)$ in \eqref{odd_trafo_01} into the standard form~\eqref{genmetric}, we are required to perform a field redefinition, leading to a new metric $\hat G$ and 
Kalb-Ramond field $\hat B$.
These steps can be represented schematically as follows:
\eq{
  \nonumber
  \arraycolsep0pt
  \begin{array}{ccccc}
  \mathcal H(G,B) & 
  \xrightarrow{\hspace{10pt}\mathcal M^t \mathcal H\hspace{1pt} \mathcal M\hspace{10pt}} &
  \widehat{\mathcal{ H}}(G,B) &
  \xrightarrow{\hspace{10pt}\hat G(G,B) \hspace{1pt},\hspace{2pt} \hat B(G,B) \hspace{10pt}} &
  {\mathcal{ H}}(\hat G,\hat B) 
  \\[4mm]
  \parbox{77pt}{\centering\scriptsize generalized metric in  variables $G$ and $B$} &
  &
  \parbox{80pt}{\centering\scriptsize $O(D,D)$-transformed generalized metric} &
  &
  \parbox{80pt}{\centering\scriptsize generalized metric in new variables} 
  \end{array}
}
Therefore,  at the level of the metric $G$ and Kalb-Ramond field $B$, the redefinitions $\hat G(G,B)$ and $\hat B(G,B)$ are the manifestation of $O(D,D)$ transformations.

In this section, we  show that for {\em every} $O(D,D)$ transformation 
of the generalized metric \eqref{genmetric}, one can read off a field redefinition for the metric $G$ and
two-form $B$. These redefinitions  take a concise form and allow for a treatment 
in terms of so-called Lie algebroids, which will be introduced in section~\ref{sec_liealg}.


\subsubsection*{Field redefinition}

Let us start by performing a general $O(D,D)$ transformation \eqref{odd_trafo_01}
on the generalized metric $\mathcal H$ 
\eq{
        \widehat{\mathcal{H}}(G,B) = \mathcal M^t\, \mathcal{H}(G,B)\, \mathcal M \,.
}
With $\mathcal M$ of the form shown in \eqref{odd_trafo_02}, 
we obtain the following expression
for the lower-right component of $\hat{\mathcal H}(G,B)$:
\eq{
  \label{comp_78}
  \widehat{\mathcal{H}}_{\rm lr} 
  = \big[d+(G-B)\,b\big]^t\, G^{-1}\,\big[d+(G-B)\,b\big]
  \,.
}
Comparing this with the original expression $\mathcal H_{\rm lr} = G^{-1}$, we see that \eqref{comp_78}
should be the inverse of the new metric $\hat G$. We therefore define
\eq{
  \label{godd}
  \widehat{G} = \gamma^{-1}\hspace{1pt} G\,(\gamma^{-1})^{t} \,,
}
where the matrix $\gamma$ is given by
\eq{
  \label{gamma}
  \gamma=d+(G-B)\,b \,.
} 
Note that, as shown in appendix \ref{app_invertible_gamma}, in the case of a Euclidean metric, i.e.\ for  $G$  positive
definite, the matrix $\gamma$ is always invertible. In particular, this includes
the most interesting case where only the internal space is described by
a non-geometric frame, whereas for the flat Minkowskian part one still
uses the geometric frame. However, to avoid confusions, we will assume the whole space-time metric to be Euclidean in the rest of this paper.
  
In order to determine the redefined
Kalb-Ramond field $\hat B$, it is convenient to consider the upper-right 
component of the generalized metric. 
In particular,  under an $O(D,D)$ transformation $\mathcal H_{\rm ur}$ transforms as
\eq{
    \widehat{\mathcal H}_{\rm ur}  = -\mathds 1 + \bigl[ c+ (G - B) a\bigr]^t
     G^{-1}  \bigl[  d+ ( G - B) b \bigr] \,.
}
After comparing with the standard form $\mathcal H_{\rm ur}=B\hspace{1pt}G^{-1}$
we are led to  the field redefinition
\eq{\label{Bodd}
	\widehat{B} = \gamma^{-1}\, \big[\gamma\,\delta^t-G\big]\, (\gamma^{-1})^{t} \,,
}
with the matrix $\delta$ defined as
\eq{
   \delta = c+(G-B)\,a \, .
}
By employing the $O(D,D)$ properties \eqref{odds}, one can show that
$\hat B$ in  \eqref{Bodd} is indeed antisymmetric. The remaining components of the generalized metric can be determined from \eqref{godd} and \eqref{Bodd} via the relation $(\eta\mathcal{H})^2=\mathds{1}$. 
To summarize, an $O(D,D)$ transformation of 
the generalized metric $\mathcal H$ gives rise to
the following field redefinitions:
\eq{
  \label{summary_fd}
  \renewcommand{\arraystretch}{1.4}
  \arraycolsep2pt
  \begin{array}{lcllcl}
    \widehat{G} &=& \gamma^{-1}\hspace{1pt} G\,(\gamma^{-1})^{t} \,, &
    \gamma&=&d+(G-B)\,b \,,
    \\
    \widehat{B} &=& \gamma^{-1}\, \big[\gamma\,\delta^t-G\big]\, (\gamma^{-1})^{t} \,, 
    \hspace{50pt}
    &
    \delta &=& c+(G-B)\,a  \,.
  \end{array}
}


\subsubsection*{Remarks}

Let us close our discussion of the field redefinitions  with the following two remarks.
First, the inverse of the relations \eqref{summary_fd} is given by 
\eq{\label{GBgb}
       G=\hat \gamma^{-1}\, \widehat G\,(\hat\gamma^{-1})^{t}
       \,,\hspace{40pt}
      B=\hat\gamma^{-1}\, \big[\,\hat\gamma\,\hat\delta^t-\widehat
        G\,\big]\, (\hat\gamma^{-1})^{t} \,,
}
written in terms of $\hat \delta$ and the inverse matrix $\gamma^{-1}=\hat{\gamma}$, which can be expressed as 
\eq{
   \label{invgamma}
   \hat\gamma=a^t+\bigl(\widehat G-\widehat B \bigr)\,b^t\,, \hspace{40pt}
   \hat\delta = c^t+\bigl(\widehat G-\widehat B\bigr)\,d^t \,.
}
Second, for the elements in the geometric subgroup $G_{\rm geom}$, the field redefinitions \eqref{summary_fd} simplify 
considerably (see also \cite{Andriot:2009fp}). In particular, for diffeomorphisms \eqref{diffeos} 
we obtain
\eq{   
    \widehat G=\mathsf A^{t}  G\hspace{1pt}\mathsf A\,,
    \hspace{50pt}
     \widehat B=\mathsf A^t B\hspace{1pt}\mathsf A\,,
 }
which is just the transformation behavior of tensors
under diffeomorphisms. 
For gauge transformations \eqref{Btrans}, given by $B$-transforms with $\mathsf B= d\Lambda$,
we also obtain
the expected transformation properties
\eq{   
    \widehat G= G \,, \hspace{40pt}
    \widehat B= B +d\Lambda\, .
    }
Since under these two types of local transformations the string effective  
action is invariant, the field redefinitions are not transcending it.
This is different for the non-geometric $\beta$-transforms,
which induce a field dependent redefinition of the metric and
the Kalb-Ramond field.
We come back to this point below.


\subsection{Examples of non-geometric frames}
\label{frames_examples}

Let us illustrate the method introduced above by two examples. 
More concretely, we
revisit two particular $O(D,D)$ transformations of the generalized metric \eqref{genmetric} which have been discussed  in the literature.


\subsubsection*{Frame I}

For the first example, we consider a setting which has recently been employed in \cite{Andriot:2011uh,Andriot:2012wx,Andriot:2012an}.
The matrix parametrizing the transformation of the generalized metric
takes the form
\eq{\label{odd1}
      \mathcal M_{I}=
    \left(\begin{matrix}  0 & (G-B\hspace{1pt}G^{-1}B)^{-1} \\
                          G-B\hspace{1pt}G^{-1}B &   0  \end{matrix}\right) ,
}
which is indeed an $O(D,D)$ transformation since the conditions \eqref{odd} are satisfied.
The transformed metric $\hat{ \mathcal H}_I(G,B)$, written in terms of the original fields $G$ and $B$, 
is then obtained as 
\eq{
  \label{ex1_genmet}
         \widehat{\cal H}_{I}  &=   \mathcal M^t_{I}\, {\cal H}\,\mathcal M_{I} \\
   &=  \left(\begin{matrix} (G-B\hspace{1pt}G^{-1}B)\hspace{1pt}G^{-1}(G-B\hspace{1pt}G^{-1}B)
   & -B\hspace{1pt}G^{-1} \\
                     G^{-1}B  & (G-B\hspace{1pt}G^{-1}B)^{-1}\end{matrix}\right) .
}
In order to express this metric  again in the form
\eqref{genmetric}, we employ the general formulas \eqref{summary_fd} to arrive at the  field redefinitions
\eq{
\label{fieldrefineA}
            \widehat G&=\hphantom{-}(\mathds{1}+B\hspace{1pt}G^{-1})\,G\,(\mathds{1}-G^{-1} B) \;,\\
            \widehat B &=-(\mathds{1}+B\hspace{1pt}G^{-1})\,B\,(\mathds{1}-G^{-1} B)\; .
}
Furthermore, it turns out to be convenient to define an antisymmetric bi-vector $\hat \beta$ 
as follows:
\eq{\label{lmubetaB}
	\hat\beta=\widehat G^{-1} \widehat B\, \widehat G^{-1}\, .
}
With the help of \eqref{lmubetaB}, we  then obtain the relation
\eq{
	(G+B)^{-1} = \widehat{G}^{-1}+\widehat{\beta}\,,
}
so that \eqref{fieldrefineA} can alternatively be written as
\eq{
  \label{new_fields_hvm}
  \arraycolsep1pt
  \begin{array}{rrcl}
      G=& (\widehat G^{-1}-\hat\beta)^{-1} & \widehat G^{-1} &
                 (\widehat G^{-1}+\hat\beta)^{-1} \,, \\[1mm]
      B=&-(\widehat G^{-1}-\hat\beta)^{-1} & \hat\beta &
                 (\widehat G^{-1}+\hat\beta)^{-1}\,,
  \end{array}                 
}
which  is precisely the field redefinition employed in \cite{Andriot:2011uh,Andriot:2012wx,Andriot:2012an}. 
Moreover, from   \eqref{fieldrefineA} we realize  that the 
$O(D,D)$ transformation \eqref{odd1} 
can also be expressed as
\eq{
  \label{comp_66-HvM}
	\mathcal M_{I}= \left(\begin{matrix}  0 & \widehat{G}^{-1} \\
		\widehat{G} &   0  \end{matrix}\right) .
}
Only for a background which is flat in the redefined variables, for instance  a toroidal one, the transformed metric is of the form $\hat G_{ab}=\delta_{ab}$.


\subsubsection*{Frame II}

The second example we want to discuss has recently
appeared in \cite{Blumenhagen:2012nk,Blumenhagen:2012nt}. 
It is characterized by an $O(D,D)$ transformation
given by the following matrix
\eq{\label{odd2}
   \mathcal M_{II}= \mathcal M_{-2B}\, \mathcal M^{\hat\beta}=
   \left(\begin{matrix}  \mathds{1} & -\hat\beta \\
                         2B  & -\mathds{1} \end{matrix}\right),
}
which consists of a combination of a $B$- and a $\beta$-transform. 
Note that in order for \eqref{odd2} to satisfy the $O(D,D)$ properties \eqref{odds},
we have to require $\hat\beta = B^{-1}$.
The  generalized metric resulting from \eqref{odd2}
is 
\eq{
  \label{genmet_03}
  \widehat{\cal H}_{II}  =  \mathcal   M_{II}^t\,    {\cal H}\, \mathcal M_{II}=
    \left(\begin{matrix} G-BG^{-1} B & -G B^{-1} \\
                B^{-1} G  & -B^{-1} G B^{-1} \end{matrix}\right)   .
}
To make a connection to \eqref{genmetric} in the standard form, we introduce 
a metric $\hat g$ on the co-tangent bundle $T^*M$ as well as
 an antisymmetric bi-vector  $\hat \beta$ by
\eq{
\label{frameBredefine}
            \hat g=-B^{-1} G\, B^{-1},\qquad
            \hat\beta=B^{-1}.
}
This field redefinition can formally be regarded  as the Seiberg-Witten limit
of \eqref{fieldrefineA}, and was studied  in detail in 
\cite{Blumenhagen:2012nk,Blumenhagen:2012nt}. 
In these variables, the transformed metric \eqref{genmet_03} is expressed as
\eq{
\widehat {\cal H}=
      \left(\begin{matrix}  \hat g^{-1}-\hat\beta^{-1}\hat g\hat\beta^{-1} &
        \hat\beta^{-1} \hat g \\
    - \hat g \hat\beta^{-1} & \hat g \end{matrix}\right)   .
}


\subsection{The quest for non-geometric actions}

In the last two subsections, we have demonstrated  how  any local $O(D,D)$ 
transformation
gives rise to  a field redefinition. 
In the following sections,  we will elaborate on the underlying structure of  the low energy
effective action of string theory expressed in terms of the redefined variables.

Recall that  the leading order action for the metric, the Kalb-Ramond field 
and the dilaton in an arbitrary number  of dimensions is\,\footnote{For matrices, $\mdet{\dots}$ denotes the absolute value of the determinant.}
\eq{
\label{stringaction}
S=-\frac{1}{2\kappa^2}
\int 
d^{n}x \hspace{1pt}\sqrt{\mdet{G}}\hspace{1pt} 
e^{-2\phi}
\Bigl(R-\tfrac{1}{12} \hspace{0.5pt} H_{abc} H^{abc}
+4 \hspace{0.25pt} \partial_a \phi \hspace{1pt}\partial^a \phi
 \Bigr) \,.
} 
This action is manifestly invariant under diffeomorphisms and under 
gauge transformations $B\to B+ d \Lambda$ of the Kalb-Ramond field, i.e.\ transformations which 
are encoded in  the geometric group $G_{\rm geom}$.
However,  upon  performing a  $\beta$-trans\-for\-ma\-tion,
the implied field redefinition is not a symmetry of the action
\eqref{stringaction}. Hence,  in the variables corresponding to a $\beta$-transform, 
the action will take a different form.

Let us illustrate this observation with  the non-geometric Frame II. 
We recall from \cite{Blumenhagen:2012nk,Blumenhagen:2012nt}  that
under the field redefinition \eqref{frameBredefine} 
the action \eqref{stringaction} changes to
\eq{
\label{finalaction}
      \hat{\mathcal S}=-\frac{1}{ 2\kappa^2} &\int d^nx\, \sqrt{\mdet{\hat g}}\: \bigl|\hat
      \beta^{-1}\bigr|\: e^{-2\phi}\, \Bigl(   \hat R - \tfrac{1}{12} \hspace{0.5pt} \widehat
      \Theta^{abc}\, \widehat\Theta_{abc}
              +4\hspace{1pt} \hat g_{ab}D^a\phi D^b \phi\Bigr)\, .
} 
Here, a new derivative operator
$D^a=\hat\beta^{am}\partial_m$ has been introduced,
$\hat R$ denotes a curvature scalar to be specified in the next section, 
and we have defined $\hat \Theta^{abc}=3 \hspace{1pt}D^{[a}
  \hat\beta^{bc]}$\,\footnote{The anti-symmetrization of $n$ indices includes  a factor $1/n!$.}.
In \cite{Blumenhagen:2012nk,Blumenhagen:2012nt} it has  been shown that the
action \eqref{finalaction} can be interpreted as coming from the differential geometry of a
{\em Lie algebroid}.
In the subsequent sections of this paper, we show  that this is just a particular
example of a more general story. Namely,   for each non-geometric
frame there exists a corresponding field redefinition 
together with a Lie algebroid, such  that  the transformed action $\hat S$ is governed
by the corresponding differential geometry.
\eq{
  \nonumber
  \fbox{\parbox{71pt}{\centering\footnotesize $O(D,D)$ transformation of gen.~metric $\mathcal H$}} 
  \hspace{4.5pt}\xrightarrow{\hspace{5pt}\eqref{summary_fd}\hspace{5pt}}\hspace{4.5pt}
  \fbox{\parbox{54pt}{\centering\footnotesize field re\-def\-i\-ni\-tions $\hat G$ and $\hat B$}} 
  \hspace{4.5pt}\xrightarrow{\hspace{5pt}{\footnotesize\rm sect.\,\ref{sec_liealg}}\hspace{5pt}}\hspace{4.5pt}
  \fbox{\parbox{58pt}{\centering\footnotesize Lie algebroid + differential geometry}} 
  \hspace{4.5pt}\xrightarrow{\hspace{5pt}{\footnotesize\rm sect.\,\ref{sec_ngdiffgeo}}\hspace{5pt}}\hspace{4.5pt}
  \fbox{\parbox{37pt}{\centering\footnotesize action $\hat S$ in a new frame}} 
}


\section{Lie algebroids}
\label{sec_liealg}

In this section, we  provide some details on the
mathematical structure of a Lie algebroid. 
Roughly speaking, a Lie algebroid is a generalization of a Lie algebra where  
the structure constants can be space-time dependent. In particular, the
Lie bracket for vector fields is generalized to a bracket for sections in a general vector bundle satisfying similar properties. 
Lie algebroids admit a natural generalization of the 
usual differential geometry framework, 
and hence covariant derivatives, torsion and curvature tensors can 
be constructed.
The relevance of Lie algebroids in the context of non-geometric fluxes 
has already been indicated in earlier work, for example in \cite{Halmagyi:2008dr,Halmagyi:2009te,Berman:2010is,Blumenhagen:2012pc}.


\subsection{Definition and examples}
\label{sec_lie_defex}

\begin{figure}[t]
\centering
\includegraphics[width=0.9\textwidth]{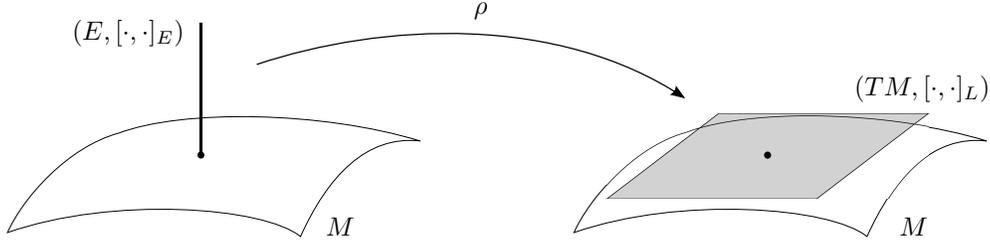}
\begin{picture}(0,0)
\put(-198,85){\footnotesize $\rho$}
\put(-254,1){\footnotesize $M$}
\put(-37,1){\footnotesize $M$}
\put(-54 ,54){\footnotesize $(TM,[\cdot,\cdot]_L)$}
\put(-350,75){\footnotesize $(E,[\cdot,\cdot]_E)$}
\end{picture}
\vskip5pt
\caption{Illustration of a Lie algebroid. On the left, one can see a manifold $M$ together with a bundle $E$ and a
bracket $[\cdot,\cdot]_E$. This structure is mapped via the anchor $\rho$ to the tangent bundle $TM$ with Lie
bracket $[\cdot,\cdot]_L$, which is shown on the right.\label{fig_01}}
\end{figure}

Let us  introduce the concept of a Lie algebroid and illustrate this structure 
by two examples.
To specify a Lie algebroid one needs three pieces of information:
\begin{itemize}

\item a vector bundle $E$ over a manifold $M$,

\item a bracket $[\cdot,\cdot ]_E : E \times E \rightarrow E$, and

\item a homomorphism $\rho : E \rightarrow TM$ called the anchor. \label{def_anchor}

\end{itemize}
A pictorial illustration for a Lie algebroid can be found in figure \ref{fig_01}.
Similar to the usual Lie bracket, we require the bracket $[\cdot,\cdot]_E$ to satisfy a Leibniz rule.
Denoting functions by $f\in {\cal C}^{\infty}(M)$ and sections  of $E$ by $s_i$, this reads
\eq{
\label{Leibniz}
[ s_1, f s_2]_E = f \hspace{1pt}[ s_1,s_2]_E + \rho(s_1)(f) s_2  \,,
} 
where $\rho(s_1)$ is a vector field which acts on $f$ as a derivation.
If  in addition the bracket $[ \cdot,\cdot]_E$ satisfies a Jacobi identity
\eq{
\label{jacobii}
\bigl[  s_1, [ s_2, s_3 ]_E \bigr]_E = \bigl[ [  s_1, s_2 ]_E , s_3 \bigr]_E +  \bigl[ s_2, [ s_1, s_3 ]_E\bigr]_E\, ,
}
then $(E,[\cdot,\cdot]_E, \rho)$ is  called
a \emph{Lie algebroid}.\footnote{If the Jacobi identity is not satisfied, the resulting structure is called a quasi-Lie algebroid.}
Therefore, in a Lie algebroid vector fields and their Lie bracket $[\cdot,\cdot]_L$ are replaced by sections of $E$ and the corresponding bracket $[\cdot,\cdot]_E$. The relation between the different brackets is established by the anchor $\rho$. Indeed, the requirement that $\rho$ is a homomorphism implies that
\eq{
\label{anhom}
	\rho\bigl( [ s_1,s_2]_E\bigr) = \bigl[\rho(s_1),\rho(s_2)\bigr]_L \,.
}

Let us illustrate this construction by two examples. The first  is the trivial example, while the
second one will be relevant in  later sections of this paper. 
\begin{itemize}

\item Consider the tangent bundle $E=TM$ with the usual Lie bracket $[ \cdot,\cdot]_E=[\cdot,\cdot]_L$. The anchor is chosen to be the identity map, i.e.\ $\rho={\rm id}$.
Then, the conditions \eqref{Leibniz} and \eqref{jacobii} reduce to the well-known properties of the Lie bracket, and \eqref{anhom} is trivially satisfied. 
Therefore, $ E=(TM,[\cdot,\cdot]_{L},\rho= \textrm{id})$ is indeed a Lie algebroid.
\label{LA_exp1}

\item As a second example, we consider a Poisson manifold $(M,\beta)$ with 
Poisson tensor $\beta =  \frac{1}{2}\,\beta^{ab}e_a \wedge e_b$, where $\{e_a\}$ denotes a
basis of vector fields.
A Lie algebroid is  given by
$E=(T^* M,[\cdot,\cdot ]_K,\rho =  \beta^\sharp)$, in which 
the anchor $\beta^{\sharp}$ is defined as
\eq{ 
\beta^{\sharp} (e^a) = \beta^{am}e_m \,,
}
with $\{e^a\}\in\Gamma(T^*M)$ the basis of one-forms dual to the vector field basis.
The bracket $[\cdot,\cdot ]_K$ on $T^*M$ is the \emph{Koszul bracket}, which for one-forms $\xi$ and $\eta$ is defined as\,\footnote{Note that for $\xi=\xi_a dx^a$ and $\eta=\eta_a dx^b$ with $\{dx^a\}$ a basis of closed one-forms, the Koszul bracket reads explicitly $[\xi,\eta]_K =\left(\xi_a \beta^{ab} \partial_b \eta_m - \eta_a \beta^{ab} \partial_b \xi_m + \xi_a \eta_b \partial_m \beta^{ab}\right) dx^m$.}
\eq{
\label{koszul}
	[\xi,\eta]_K = L_{\beta^\sharp(\xi)}\eta
 -\iota_{\beta^\sharp(\eta)}\,d\xi \, ,
}
where the Lie derivative on forms is given by $L_{X} = \iota_X \circ d + d \circ \iota_X$ with $d$ the de Rham differential. The conditions \eqref{Leibniz}, \eqref{jacobii} and \eqref{anhom} are satisfied, provided that $\beta$ is a Poisson tensor, i.e.\ $\beta^{[a|m}\partial_m \beta^{|bc]}=0$. 

\end{itemize}


\subsection{Differential  geometry of a Lie algebroid}
\label{sec_alggeom}

After having introduced the concept of a Lie algebroid, we now turn to the corresponding
differential geometry. We will be brief here, but  more details 
can be found in \cite{2008arXiv0806.3522B}. 
To get a general idea about the construction, let us note that the standard Riemann curvature 
tensor is based on the Lie bracket. Hence, a natural generalization to Lie algebroids is given by 
replacing the Lie bracket as $[\cdot, \cdot]_L \to [\cdot, \cdot]_E$ and
inserting the anchor $\rho$ whenever needed. This can be regarded as 
the main guiding principle  for the following.


\subsubsection*{Covariant derivative}

Let us start our discussion by defining a partial derivative. With $s\in\Gamma(E)$ a section of the 
bundle $E$ and $f\in{\cal C}^{\infty}(M)$  a function, we define
\eq{
      D_s f= \rho(s)\hspace{1pt} f\, .
}
For our two examples on page \pageref{LA_exp1} above, this means the following:
\eq{
  \arraycolsep2pt
  \begin{array}{lc@{\hspace{15pt}}lcl@{\hspace{30pt}}l@{\hspace{5pt}}cl}
  E= TM &:& D_{e_a} f &=& \partial_a f& {\rm where} &s = e_a & \mbox{is a basis vector field,} \\[2mm]
  E= T^*M &:& D_{e^a} f &=& \beta^{am}\partial_m f& {\rm where} &s = e^a & \mbox{is a basis one-form.} 
  \end{array}
}

Concerning  the covariant derivative, we recall  that in the usual case $\nabla$ takes two vector fields and 
assigns to them a third one. This generalizes  to a map 
$\widehat{\nabla}:\Gamma(E)\times\Gamma(E) \to\Gamma(E)$ 
which satisfies the following three properties
\eq{
  \label{def_covder}
  &\widehat\nabla_{s_1}(s_2+s_3) = \widehat\nabla_{s_1} s_2 + \widehat\nabla_{s_1} s_3 \,, \\[1mm]
  &\widehat\nabla_{s_1}(fs_2)= f\,\widehat\nabla_{s_1}(s_2) + \rho(s_1)f\cdot s_2 \,, \\[1mm]
  &\widehat\nabla_{(fs_1)} s_2 = f\,\widehat\nabla_{s_1}s_2 \,,
}
for  functions $f\in C^\infty(M)$ and section $s_i \in \Gamma(E)$.
The extension to tensors of higher degree is obtained via the Leibniz rule.
The action of the covariant derivative on sections $t^*\in \Gamma(E^*)$ of the dual bundle $E^*$ is
determined via the  compatibility with the insertion 
 $\langle\cdot,\cdot\rangle$.\footnote{
 The insertion $\langle \cdot,\cdot \rangle : E^* \times E \to \mathbb R$
 is characterized by $\langle \epsilon^{\alpha}, \epsilon_{\beta} \rangle = \delta^{\alpha}{}_{\beta}$ for $\{\epsilon_{\alpha}\}\in\Gamma(E)$ a basis of $E$
 and $\{\epsilon^{\alpha}\}\in \Gamma(E^*)$ the corresponding dual basis. 
 }
We have
 \eq{	
 \widehat\nabla_{s_1} \langle t^* , s_2 \rangle = 
 \rho(s_1)\langle t^* , s_2 \rangle  = \langle\widehat\nabla_{s_1} t^*,s_2\rangle
  + \langle t^* ,\widehat\nabla_{s_1} s_2\rangle \,.
}
Introducing a local frame $\{\epsilon_{\alpha}\}$ for $E$ and its dual $\{\epsilon^{\alpha}\}$, we define the Christoffel symbols by
$\widehat{\Gamma}^{\gamma}{}_{\alpha\beta} = \iota_{\epsilon^{\gamma}}\widehat{\nabla}_{\epsilon_{\alpha}}\epsilon_{\beta}$.
Using then \eqref{def_covder}, we can  write locally
\eq{
	\widehat{\nabla}_{\epsilon^{\alpha}}s^{\beta} = D_{\alpha} s^{\beta}
		+ \widehat{\Gamma}^{\beta}{}_{\alpha\gamma}\,s^{\gamma} \hspace{40pt}
		{\rm for} \hspace{20pt} s = s^{\alpha} \epsilon_{\alpha} \,.
}
Let us emphasize that this construction is in complete analogy with the standard differential geometry calculus.
We only employed a more general bundle and inserted the anchor map $\rho$ when
needed.


\subsubsection*{Curvature and torsion tensors}

After having defined a covariant derivative, we can  define curvature and torsion
tensors. This is again in analogy to the standard case. For the curvature tensor we write
\eq{
  \label{T&R}
  \widehat R(s_1,s_2)\, s_3 = \bigl[\widehat \nabla_{s_1},\widehat \nabla_{s_2}\bigr]\, s_3 
  - \widehat \nabla_{[ s_1,s_2]_E}\, s_3 \,,
}            
where $s_i\in\Gamma(E)$ are sections of $E$. Note that when replacing $s_i$ by vector fields $X,Y,Z$ 
and $[\cdot ,\cdot]_E$ by the Lie bracket, we recover the familiar definition of the  Riemann curvature tensor.
For the torsion tensor we have similarly
\eq{
  \label{torsion}
  \widehat T(s_1,s_2) =\widehat \nabla_{s_1} s_2 -\widehat \nabla_{s_2} s_1-[ s_1,s_2]_E \,.
}
To show that these expressions are indeed tensors
with respect to diffeomorphisms, one has to check that they
are $\mathcal C^\infty(M)$-linear in all arguments. In case of, for instance,
  the torsion tensor, this means 
\eq{
  \widehat T(f\hspace{1pt} s_1, g\hspace{1pt} s_2) = f\hspace{1pt}g\, \widehat T(s_1,s_2) \,,
}
for functions $f,g\in\mathcal C^{\infty}(M)$,
which can be  checked explicitly using \eqref{def_covder} as well as 
the Leibniz property \eqref{Leibniz}.


\subsubsection*{Metric and Levi-Civita connection}
\label{LA_metric}

Let us finally introduce a metric $\fa g$ on the Lie algebroid $(E,[\cdot,\cdot]_E, \rho)$,
which  is an element in $\Gamma(E^*\otimes_{\rm sym}\hspace{-1pt}E^*)$
 assigning a number to a pair of sections $s_1,s_2\in \Gamma(E)$. 
In the case of our first example on page  \pageref{LA_exp1} this reads 
\eq{
  G( X, Y ) =  X^a G_{ab} Y^b \,,
}
for $G= G_{ab} \hspace{1pt} dx^a \otimes_{\rm sym} dx^b$ and vector fields $X=X^a \partial_a$ and $Y= Y^b \partial_b$.
We require the metric $\fa g$ to be compatible with
the connection, which means that for sections $s_i\in \Gamma(E)$
\eq{
 \widehat \nabla_{s_1}\big(\fa g( s_2 , s_3 )\big) = 
 \fa g\bigl(\widehat \nabla_{s_1} s_2,s_3 \bigr)
  + \fa g\bigl( s_2 ,\widehat \nabla_{s_1} s_3 \bigr) \,.
}
If we demand in addition that the torsion tensor \eqref{torsion} vanishes, 
then  a particular covariant derivative, the so-called
Levi-Civita connection, is uniquely determined.
The latter is given by the  Koszul formula 
\eq{
  \label{koszul_formula}
  &\fa g\bigl(\widehat \nabla_{s_1}s_2,s_3\bigr) =
  \tfrac{1}{2}\Big[ \rho(s_1)\,\fa g(s_2,s_3)+\rho(s_2)\,\fa g(s_3,s_1)-\rho(s_3)\,\fa g(s_1,s_2)\\	
  & \hspace{98pt}
  +\fa g([ s_1,s_2]_E,s_3)+\fa g([ s_3,s_1]_E,s_2)-\fa g([ s_2,s_3]_E,s_1)\Big] \,.
}
In the following, the connection $\widehat \nabla$ is always understood to be Levi-Civita.

After having introduced the general theory, we will now give two equivalent constructions for Lie algebroids suitable for describing the field redefinitions \eqref{summary_fd} geometrically.


\subsection{Lie algebroids on $TM$}
In section~\ref{sec_oddredef} we have derived the field redefinitions
\eqref{summary_fd} associated to an $O(D,D)$ transformation. Interestingly,
 the metric  transforms by conjugation with the matrix
$\gamma=d+(G-B)\,b$. In this section, we   deduce an anchor map 
together with an associated
bracket from $\gamma$, thus yielding a Lie algebroid for every field 
redefinition.


\subsubsection*{Identifying an anchor}

Let us start by considering a Lie algebroid on the tangent bundle $E=TM$ of a manifold $M$, 
where the anchor map is related to the matrix $\gamma$.
Recalling  the submatrices $a,b,c$ and $d$ in a general $O(D,D)$ transformation \eqref{odd_trafo_02}, and keeping in mind the index structure displayed in \eqref{index_structure},
we have the following linear mappings
\eq{	
  \arraycolsep2pt
  \begin{array}{r@{\hspace{6pt}}lcl@{\hspace{50pt}}r@{\hspace{6pt}}lcl}
  a: & TM &\to& TM\,, &
  b: & T^*M &\to& TM \,, 
  \\[2mm]
  c: & TM &\to& T^*M \,, &
  d: & T^*M & \to & T^*M \,.
  \end{array}
}
Furthermore, the matrix $(G-B)$ can be considered as  $(G-B):TM\to T^*M$ so
that we obtain
\eq{
  \label{gamma_04}
  \gamma :\hspace{6pt} T^*M\to T^*M \,.
}
Our aim is  to identify an anchor $\rho: E \to TM$ which maps elements of the Lie algebroid bundle $E=TM$ 
to the tangent bundle $TM$. A natural candidate is \eqref{gamma_04}, defined on the dual spaces. To determine the anchor, note that for a linear map $f: V\to W$ we have
\eq{\label{dualsec}
  \arraycolsep2pt
  \begin{array}{rcc@{\hspace{10pt}}lclcl}
  &f &:& V & \to & W \,,&& \\
  &f^{-1} &:& W & \to & V \,, && \\ 
  &f^t &:& W^* & \to & V^* \;,&\hspace{10pt}&\omega\mapsto \omega\circ f \,, \\ 
  f^*=&(f^t)^{-1} &:& V^* & \to & W^* \,,&&\nu\mapsto\nu\circ f^{-1} \,.
  \end{array}
}
Recalling \eqref{godd}, $\gamma$ has to be considered as  a map $E^*\to T^*M$. Therefore, the  anchor $\rho : TM \to TM$ following from \eqref{gamma_04} is given by the inverse-transpose of $\gamma$
\eq{
  \label{anchor}
  \rho=(\gamma^{-1})^t \;.
}


\subsubsection*{Lie algebroid bracket}

Let us now determine a bracket for the Lie algebroid bundle $E=TM$. One of the
main requirements on $[\cdot,\cdot]_E$ is that the anchor \eqref{anchor} is a homomorphism, which means   $\rho$ has to satisfy equation \eqref{anhom}. 
We start by noting that for a vector field $X=X^a e_a$ we have
\eq{
  \rho (X)= (\rho^a{}_b\, X^b)\, e_a = X^a (\rho^t)_a{}^b  e_b = X^a D_a \,,
}
where we defined the partial derivative for the Lie algebroid as
\eq{	
   D_a = (\rho^t)_a{}^b\,  e\vphantom{(\rho^t)}_b \,.
}
In general, $\{e_a\}=\{\partial_a\}$ is a non-holonomic basis of $TM$ which for the Lie bracket implies  $[ e_a, e_b]_L = f_{ab}{}^c\hspace{1pt} e_c$ with $f_{ab}{}^c$ the structure constants of the underlying Lie algebra.
For two vector fields $X=X^ae_a$ and $Y=Y^b e_b$ we then compute
\eq{
  \label{HvM_9}
  \bigl[ \rho(X), \rho(Y) \bigr]_L = 
 \big(X^mD_mY^a-Y^mD_mX^a +X^m\,Y^n\,F_{mn}{}^a\big)\,(\rho^t)_a{}^b\, e_b \,,
}
where we have defined
\eq{
  \label{struct_const}
  F_{ab}{}^c =
  (\rho^{-1})^c{}_m
  \bigl[ D_a (\rho^t)_b{}^m - D_b (\rho^t)_a{}^m + (\rho^t)_a{}^p\,
    (\rho^t)_b{}^q \, f_{pq}{}^m \bigr]
  \,.
}
This suggests to define a new bracket $\lb \cdot,\cdot \rb$ on $E=TM$ of the following form
\eq{\label{tbracket}
  \lb X,Y\rb = \big(X^mD_mY^a-Y^mD_mX^a +X^m\,Y^n\,F_{mn}{}^a\big)\, e_a \,.
}
Indeed, noting that $\rho(e_a) = (\rho^t)_a{}^b e_b$ and comparing with \eqref{HvM_9}, 
we see that this bracket  satisfies the homomorphism property \eqref{anhom}
\eq{
  \rho\big( \lb  X,Y\rb\big)=[\rho (X),\rho (Y)]_L \,.
}
Furthermore, by construction the new bracket $\lb\cdot,\cdot\rb$ satisfies the 
Jacobi identity \eqref{jacobii} as well as the
Leibniz rule \eqref{Leibniz}
\eq{
	\lb X,fY \rb = f \hspace{1pt} \lb
  X,Y\rb + (X^aD_af) \hspace{1pt}Y\,.
}


\subsubsection*{Remark}

In the previous paragraph, we have shown that for every $O(D,D)$ transformation
we can construct a corresponding  Lie algebroid $(TM,\lb\cdot,\cdot\rb,\rho)$ on the tangent 
bundle $TM$. 
However, one may argue that such a Lie algebroid 
could also be obtained by describing  $TM$ in 
a particular non-holonomic basis.

Indeed, let us define a basis $\{ \tilde e_a\}$ of vector fields as $\tilde e_a = (\rho^t)_a{}^b\,  e_b{}^i\, \partial_i$, where
$\{\partial_i\}\in\Gamma(TM)$ is a holonomic basis with
$[\partial_i,\partial_j]_L=0$.
For the Lie bracket in this basis we then find $[\tilde e_a,\tilde e_b]_L = F_{ab}{}^c\, \tilde e_c$, or 
equivalently
\eq{
	\lb X,Y \rb = [X^a \tilde e_a,Y^b\tilde e_b]_L \,.
}
Or in other words, the Lie algebroid bracket $\lb \cdot, \cdot \rb$ is just the ordinary Lie bracket in the basis
$\{ \tilde e_a\}$.
Therefore, for any anchor $\rho=(\gamma^t)^{-1}$ we could choose a corresponding 
diffeomorphism $\gamma = (\mathsf{A}^t)^{-1}$ which gives rise to a
Lie algebroid bracket.
In the case of geometric transformations $\mathcal{M}\in G_{\mathrm{geom}}$, 
this is the expected form, but for $\beta$-transforms \eqref{betatrans} with $\gamma=1+(G-B)\beta$
the corresponding diffeomorphism $(\mathsf{A}^t)^{-1}=1+(G-B)\beta$
involves the dynamical fields
themselves. 
This is not what one usually understands by a 
diffeomorphism in differential geometry, and
must rather be considered as a
\emph{generalized} change of coordinates.

These observations can
be summarized by saying that $\beta$-transforms go beyond the
usual notions of differential geometry, and the Lie algebroid presented
in this section provides the appropriate mathematical framework
to describe both geometric transformations  $G_{\mathrm{geom}}$ and 
non-geometric $\beta$-transforms.


\subsection{Lie algebroids on $T^*M$}
\label{sec_tstar}

After having  constructed a Lie algebroid on $TM$, we next
investigate how a Lie algebroid structure can be defined 
on the cotangent bundle $T^*M$. 
For our second example in section \ref{frames_examples}, such a Lie algebroid was constructed 
in \cite{Blumenhagen:2012nt}.


\subsubsection*{Construction}

Let us  note that the metric $G$ on the manifold $M$ can be seen as a linear mapping $G: TM \to T^*M$, while the inverse gives a map $G^{-1}:T^*M\to TM$. Combining this observation with 
\eqref{gamma_04}, we arrive at the following picture
\begin{equation}
\label{diagram}
\begin{CD}
E_{2} = T^*M     @>\hspace{27pt}\gamma\hspace{27pt}>>  T^*M\\
@V\hat G^{-1}VV        @VVG^{-1}V\\
E_{1} = TM     @>>\quad\rho= (\gamma^t)^{-1}\quad>  TM
\end{CD}
\end{equation}
where on the left-hand side we have the Lie algebroid bundles $E_1=TM$ and $E_2=T^*M$, while on
the right-hand side there are the standard tangent and cotangent bundles of the manifold.
An anchor for a Lie algebroid on $T^*M$ can therefore be defined as follows
\eq{\label{danchor}
	\tilde{\rho}=  G^{-1}\circ\gamma:\hspace{10pt}T^*M\to TM \,.
}
For a one-form $\xi= \xi_{\alpha} e^{\alpha}$, locally the anchor $\tilde\rho$ acts as follows
\eq{
	\tilde{\rho}\hspace{1pt}(\xi) =
	\bigl( \tilde \rho^{b\alpha}\xi_{\alpha}\bigr) e_b = 
	\xi_\alpha\,(\tilde{\rho}^t)^{\alpha b}\hspace{1pt}e_b 
	= \xi_\alpha\,(\gamma^t)^\alpha{}_{\beta}\, \hspace{1pt}G^{\beta c}\,e_c \,,
}
where we denote  indices related to $T^*M$ by Greek letters.
Analogous to the bracket \eqref{tbracket} on $TM$, we can define a bracket on $T^*M$ as
\eq{\label{dtbracket}
	\lb\hspace{1pt}\xi,\eta\hspace{1pt}\rb_* = \big(\xi_\mu\hspace{1pt}D^\mu\eta_\alpha
	-\eta_\mu\hspace{1pt}D^\mu\xi_\alpha
		+\xi_\mu\,\eta_\nu\,Q_\alpha{}^{\mu\nu}\big)e^\alpha \,,
}
with the associated partial derivative given by 
\eq{
	D^\alpha=(\tilde{\rho}^t)^{\alpha m}e_m \,,
}
and structure constants of the form
\eq{
\label{Q}
  Q_\alpha{}^{\beta\gamma}=
  (\tilde\rho^{-1})_{\alpha m} 
  \big[ D^\beta(\tilde{\rho}^t)^{\gamma m}-D^{\gamma}(\tilde{\rho}^t)^{\beta m}
  + (\tilde\rho^t)^{\beta p}(\tilde\rho^t)^{\gamma q} f_{pq}{}^m
  \big] \,.
}
Again, one can verify that  \eqref{dtbracket} satisfies the homomorphism 
property \eqref{anhom} as well as  the corresponding Leibniz rule and Jacobi identity. 
Therefore,  we obtain a Lie algebroid
$(T^*M,\lb \cdot,\cdot \rb,\tilde{\rho})$ on the cotangent bundle.


\subsubsection*{Remarks}

Let us close this subsection with two remarks:
\begin{itemize}

\item For an antisymmetric anchor with an
appropriate Poisson condition, the bracket \eqref{dtbracket} coincides with the
corresponding Koszul bracket shown in equation \eqref{koszul}
(cf.\ \cite{Gualtieri:2003dx}).
This is the realm of Poisson geometry. However,
\eqref{dtbracket} is more general in the sense that it is also valid
for the symmetric part of an anchor.

\item The bracket \eqref{dtbracket} on the cotangent bundle $T^*M$ can be related to the 
bracket \eqref{tbracket} on $TM$ via 
\eq{
	\bigl\lb\hspace{1pt}\xi,\eta\hspace{1pt}\bigr\rb_{*}
	= \widehat{G}\Big(\big\lb\widehat{G}^{-1}\xi,\widehat{G}^{-1}\eta\big\rb
		\Big)\,,
}
where $\widehat{G}$ is the transformed metric \eqref{godd}.
Thus, with the metric only the indices are raised and lowered, which 
means that the differential geometry constructed on
$(T^*M,\lb\cdot,\cdot\rb_*,\tilde{\rho})$ is equivalent to the 
one constructed on $(TM,\lb\cdot,\cdot\rb,\rho)$.

\end{itemize}


\subsection{Examples}

Let us illustrate the above constructions  within the two frames mentioned in section~\ref{frames_examples}. More concretely, we  determine explicitly the  Lie algebroids corresponding to the $O(D,D)$ transformations \eqref{odd1} and \eqref{odd2}.


\subsubsection*{Frame I}

Inserting the $O(D,D)$ transformation \eqref{odd1} into the map 
\eqref{gamma} yields the matrix
\eq{
	\gamma_\mathrm{I} = (\mathds{1}+BG^{-1})^{-1}\,.
}
Together with \eqref{invgamma}, we can then confirm that the general formulas \eqref{godd} 
and \eqref{Bodd} reproduce the field redefinition \eqref{fieldrefineA}. The anchor 
\eqref{anchor} 
is given by
\eq{\label{anchorI}
	\rho_\mathrm{I} = \mathds 1-G^{-1}B \,,
}
and the corresponding structure constants  of the Lie algebroid bracket $\lb \cdot,\cdot\rb_{\rm I}$
can be computed from \eqref{struct_const}. 
In particular, we find
\eq{
  \label{ex_6104}
  (F_I)_{ab}{}^c = 2\,
  \bigl[ (G+ B)\hspace{1pt} G^{-1} \bigr]_{[a}^{\hspace{9pt}m}\,
  \partial_m \bigl[ B G^{-1} \bigr]_{b]}^{\hspace{9pt}n}
  \bigl[ G\hspace{1pt}(G+B)^{-1} \bigr]_n^{\hspace{5pt} c}
  \,,
}
where for simplicity we have set to zero the structure constants $f_{ab}{}^c$ of the coordinate indices.
Working out in detail \eqref{ex_6104} results in a rather lengthy expression which we 
do not present here. 
However, the above information completely characterizes the Lie algebroid 
$(TM,\lb\cdot,\cdot\rb_\mathrm{I},\rho_\mathrm{I})$.


\subsubsection*{Frame II}

The transformation \eqref{odd2} can be used to provide an example of a Lie 
algebroid on the cotangent bundle. From the $O(D,D)$ transformation 
we can read off the map 
\eq{
  \gamma_\mathrm{II} = - \mathds 1 -(G-B)\hat{\beta} = -G\,\hat{\beta} \,,
}
where we employed $\hat{\beta}=B^{-1}$. Again, using \eqref{godd} and
\eqref{Bodd} we can confirm the redefinition \eqref{frameBredefine}. In
addition, invoking  \eqref{anchor} and \eqref{danchor} we obtain the
corresponding anchor on $TM$ and $T^*M$ 
as
\eq{\label{anchorII}
	\rho_\mathrm{II} = G^{-1}\hat{\beta}^{-1}  \hspace{30pt}\mbox{and}\hspace{30pt}
	\tilde{\rho}_\mathrm{II} = -\hat{\beta} \,,
}
respectively. The structure constants \eqref{struct_const} and \eqref{Q} of the Lie algebroid 
brackets on $TM$ and $T^*M$ are computed as follows,
\eq{
  (F_{II})_{ab}{}^c &= 2\,
  \bigl[ \hat\beta^{-1}\hspace{1pt} G^{-1} \bigr]_{[a}^{\hspace{9pt}m}\,
  \partial_m \bigl[ \hat\beta^{-1}\hspace{1pt} G^{-1} \bigr]_{b]}^{\hspace{9pt}n}
  \bigl[ G\hspace{1pt}\hat\beta \bigr]_n^{\hspace{5pt} c}
  \,, \\
  (Q_{II})_{\alpha}{}^{\beta\gamma} &= 2\,
    \hat\beta^{\,[ \beta |m}\,
  \partial_m \hat\beta^{\, |\gamma] n}
  [ \hat\beta^{-1} ]_{n\alpha}
  \,,  
}
where for simplicity we set  $f_{ab}{}^c$ to zero. Note that the structure constants $Q_{II}$ take
a particular simple form for this example and match with the corresponding 
expression in \cite{Blumenhagen:2012nt}. 
Furthermore, in view of our observations at the end of section~\ref{sec_tstar}, the anchor
$\tilde{\rho}_\mathrm{II}$ is interesting as it is
antisymmetric. 
If we require $\hat\beta$ to satisfy the
\emph{quasi-Poisson} 
condition
\eq{
	\hat{\beta}^{am}\partial_m\hat{\beta}^{bc} + \mathrm{cycl.} 
	= -\hat{\beta}^{am}\,\hat{\beta}^{bn}\,\hat{\beta}^{ck}\,H_{mnk}
}
for a three-form $H=\tfrac{1}{3!} H_{\alpha\beta\gamma} \hspace{1pt}e^{\alpha}\wedge e^{\beta}\wedge e^{\gamma}$, the bracket \eqref{dtbracket} coincides with the so-called $H$-twisted Koszul bracket.
Indeed, we find 
\eq{
	\bigl\lb \hspace{1pt}\xi,\eta\hspace{1pt}\bigr\rb_\mathrm{II} 
	= L_{\hat{\beta}(\xi)}\eta - \iota_{\hat{\beta}(\eta)}d\xi 
		+ \iota_{\hat{\beta}(\eta)}\,\iota_{\hat{\beta}(\xi)}\,H
	= \bigl[\hspace{1pt}\xi,\eta\hspace{1pt}\bigr]_K^H \,,
}
where as before $\hat\beta(\xi)= \xi_{\alpha} \hat\beta^{\alpha b} e_b$
and $L_{X} = \iota_X \circ d + d \circ \iota_X$.
This 
provides the connection to \cite{Blumenhagen:2012nt} where this particular Lie algebroid $(T^*M,[\cdot,\cdot]_K^H,\hat{\beta})$ has been studied in
detail.


\section{Differential geometry in non-geometric frames}
\label{sec_ngdiffgeo}

In this section, we establish a connection between the differential geometry 
of a Lie algebroid on $E$, on the one hand, 
and the standard geometry on $TM$, on the other hand.
In particular, utilizing the field redefinitions \eqref{summary_fd},
we derive a  correspondence between the respective differential 
geometric objects.
This provides a general framework for the formulation of  gravity 
theories which are related to standard gravity 
via $O(D,D)$ transformations.

Our setup is as follows: we start from a general 
Lie algebroid $(E,[\cdot,\cdot]_E,\rho)$ 
equipped with a metric 
$g\in\Gamma(E^*\otimes_{\mathrm{symm}}E^*)$ and  
for which the anchor $\rho:E\to TM$ 
is invertible.
For our previous example of a Lie algebroid on $TM$ one has  $g=\widehat G$.
Moreover, we assume this metric to be related to the Riemannian metric 
$G$ by applying the anchor as follows
\eq{\label{genGg}
	G = \big(\!\otimes^2\!\rho^*\big)(g) \qquad\Longleftrightarrow\qquad
	g = \big(\!\otimes^2\!\rho^t\big)(G) \,,
}
where $\rho^*:E^*\to T^*M$ is the dual anchor and $\rho^t:T^*M\to E^*$ the 
transpose anchor, cf.\ \eqref{dualsec}.
The relation \eqref{genGg} contains the redefinition discussed above
as it is in accordance with \eqref{godd} for
$\rho=(\gamma^{-1})^t$.


\subsection{Relating Riemannian geometry to non-geometry}

In this section we work out in detail the relation between the
differential geometric objects appearing for the Lie algebroid
and the familiar ones  from standard Riemannian geometry.\footnote{The reader not interested in the mathematical details may go directly to page \pageref{geom_summary}, where a summary of all relevant formulas of this subsection can be found.}
Let $\{e_a\}$ and $\{\epsilon_\alpha\}$ be a local frame for $TM$ and $E$,
respectively. Using the corresponding dual bases, we can write the metrics 
as $G=G_{ab}\, e^a\otimes e^b$  and
$g=g_{\alpha\beta}\, \epsilon^\alpha\otimes\epsilon^\beta$. 
Thus, the field redefinition \eqref{genGg} in local coordinates reads\,\footnote{The conventions for the indices are as follows
\eq{
	\rho \equiv \rho^a{}_\alpha\,, \hspace{20pt}
	\rho^{-1}\equiv (\rho^{-1})^\alpha{}_a \,, \hspace{20pt}
	\rho^t \equiv (\rho^t)_\alpha{}^a\,,\hspace{20pt}
	\rho^*\equiv (\rho^*)_a{}^\alpha \,.
}
Note that here the index $\alpha$ of $\rho$, i.e.\ the one 
corresponding to the Lie algebroid,  is chosen to be downstairs. However, 
when discussing particular examples, for instance $E=T^*M$, it might be more convenient to change the index structure to $\rho^{a \beta}$.
}
\eq{\label{gGloc}	
G_{ab} =
  (\rho^*)_a{}^{\alpha}(\rho^*)_b{}^{\beta}\,g_{\alpha\beta}\,.
}
In a coordinate-free notation,
for sections $s,t\in\Gamma(E)$ one  can equivalently write
\eq{\label{gG}
	G(\rho(s),\rho(t))= g(s,t) \;,
}
as $\rho^*=(\rho^t)^{-1}$ and for a one-form $\xi \in T^*M$ one has
\eq{
	\rho^t(\xi)(s) = \big(\xi_a\, (\rho^t)_\alpha{}^a\big)\, s^\alpha 
	= \xi_a\, \big(\rho^a{}_\alpha\, s^\alpha\big) = \xi(\rho(s)) \,.
}
In the following, sections of $E$ are denoted by $s,t$ and dual sections by 
$s^*,t^*$.


\subsubsection*{The connections}

Let us turn to the Levi-Civita connection on the Lie algebroid $E$.
Denoting the standard Levi-Civita connection on $TM$ by $\nabla$
and employing \eqref{gG} in the Koszul formula \eqref{koszul_formula} together with the anchor property \eqref{anhom}, we find
\eq{
	G\big(\rho(\widehat{\nabla}_rs),\rho(t)\big) &= g\big(\widehat{\nabla}_rs,t\big) \\
	&= G\big(\nabla_{\rho(r)}\rho(s),\rho(t)\big) \,.
}
Thus, by non-degeneracy of the metrics we infer
\eq{\label{connrel}
	\rho(\widehat{\nabla}_st) = \nabla_{\rho(s)}\rho(t) \,,\hspace{50pt}
	\rho^*(\widehat{\nabla}_s t^*) = \nabla_{\rho(s)}\rho^*(t^*) \,.
}
The second identity follows from compatibility with the insertion and the first identity. This can be seen as follows. First observe that $\langle s,t^*\rangle = \langle\rho(s),\rho^*(t^*)\rangle$. In view of the compatibility  of $\widehat{\nabla}$ and $\nabla$ with the insertion, this implies
\eq{
	\langle\widehat{\nabla}_rs,t^*\rangle + \langle s,\widehat{\nabla}_r t^*\rangle
	&= \langle\nabla_{\rho(r)}\rho(s),\rho^*(t^*)\rangle
		+ \langle\rho(s),\nabla_{\rho(r)}\rho^*(t^*)\rangle \\
	&= \langle\widehat{\nabla}_rs,t^*\rangle 
		+ \langle\rho(s),\nabla_{\rho(r)}\rho^*(t^*)\rangle \\
	&= \langle\widehat{\nabla}_rs,t^*\rangle 
		+ \langle s,\rho^t(\nabla_{\rho(r)}\rho^*(t^*))\rangle\,.
}
We therefore have
\eq{
	\widehat{\nabla}_r t^* = \rho^t(\nabla_{\rho(r)}\rho^*(t^*))
	\qquad\Longleftrightarrow\qquad
	\rho^*(\widehat{\nabla}_s t^*) = \nabla_{\rho(s)}\rho^*(t^*) \,,
}
and so \eqref{connrel} establishes the connection between the Levi-Civita connections in both frames.
The corresponding connection coefficients in local coordinates 
are defined in the standard way
\eq{
	\Gamma^c{}_{ab} = \iota_{e^c}\,\nabla_{e_a}e_b \,,\hspace{50pt}
	\widehat{\Gamma}^\gamma{}_{\alpha\beta} 
	= \iota_{\epsilon^\gamma}\,\widehat{\nabla}_{\epsilon_\alpha}\epsilon_\beta \,.
}
Using then \eqref{connrel},  the relation between the Christoffel symbols in the Riemannian and
Lie algebroid setting reads
\eq{\label{Christoffel}
	\widehat{\Gamma}^\gamma{}_{\alpha\beta}= 
		(\rho^{-1})^{\gamma}{}_c\,\rho^a{}_\alpha\,\rho^b{}_\beta\,\Gamma^c{}_{ab}
		+(\rho^{-1})^\gamma{}_b\,\rho^a{}_\alpha\,\partial_a \rho^b{}_\beta\,.
}


\subsubsection*{Torsion and curvature}
The relation \eqref{connrel} found above is of the same type as the relation
between the brackets given by the anchor property, i.e.\
$\rho([s,t]_E)=[\rho(s),\rho(t)]_L$. Since the 
torsion and the curvature are defined in terms of the connection and the bracket
(cf.\ \eqref{T&R} and \eqref{torsion}), we can relate them accordingly. Thus, by
applying \eqref{connrel} and the anchor property \eqref{anhom} we 
obtain
\eq{
	\widehat{T}(s,t) = \rho^{-1}\big(T(\rho(s),\rho(t))\big) \,, \hspace{40pt}
	\widehat{R}(s,t)r = \rho^{-1}\big(R(\rho(s),\rho(t))\rho(r)\big) \,,
}
where $T$ and $R$ denote the torsion and curvature with respect to $\nabla$ on $TM$. 
In a local frame, the relation between the curvatures reads
\eq{
\label{curvrelloc}
 	\widehat{R}^\alpha{}_{\beta\gamma\delta} =  \langle\epsilon^\alpha,
		\widehat{R}(\epsilon_\gamma,\epsilon_\delta)\epsilon_\beta\rangle
	=  (\rho^{-1})^\alpha{}_a\,\rho^b{}_\beta\,\rho^c{}_\gamma\,\rho^d{}_\delta\,
		R^a{}_{bcd} \,,
}
which is simply the contraction of all indices of the Riemann tensor $R^a{}_{bcd}$ with the anchor.
For the Ricci tensor and Ricci scalar this implies
\eq{
	\widehat{R}_{\alpha\beta} = \widehat{R}^\gamma{}_{\alpha\gamma\beta}
	= \rho^a{}_\alpha\,\rho^b{}_\beta\, R_{ab} \,,
	\hspace{40pt}
	\widehat{R} = g^{\alpha\beta}\widehat{R}_{\alpha\beta} = G^{ab}\,R_{ab} = R \,,
}
where we employed \eqref{gGloc} for the Ricci scalar. 
Let us remark that for a
covariant theory, all terms appearing in the corresponding Lagrangian must be
scalars. From \eqref{curvrelloc} and \eqref{gGloc} we then infer that all
scalars built from curvature tensors are equal, e.g.\  $\widehat{R}_{\alpha\beta}\widehat{R}^{\alpha\beta}=R_{ab}R^{ab}$.


\subsubsection*{The exterior derivative}

As was done for the connection, also the exterior derivative can be transferred to the Lie 
algebroid by applying the anchor.
Indeed, any Lie algebroid can be 
equipped with a nilpotent exterior derivative as follows
\eq{\label{algd}
	d_E\, \theta^*(s_0,\dots,s_n) =&\sum_{i=0}^n(-1)^i\,\rho(s_i)\,
		\theta^*(s_0,\dots,\hat{s_i},\dots,s_n) \\
		&+\sum_{i<j}(-1)^{i+j}\,
		\theta^*([s_i,s_j]_E,s_0,\dots,\hat{s_i},\dots,\hat{s_j},\dots,s_n)\,,
}
where $\theta^*\in\Gamma(\Lambda^nE^*)$ is the analog of an 
$n$-form on the Lie algebroid and $\hat s_i$ indicates the omission of that entry.
The Jacobi identity of the bracket $[\cdot,\cdot]_E$ implies
that \eqref{algd} satisfies $(d_E)^2=0$. The anchor property and the corresponding 
formula for the de~Rahm differential allow to compute
\eq{\label{diffs}
	\Big(\big(\Lambda^{n+1}\!\rho^*\big)(d_E\,\theta^*)\Big)(X_0,\dots,X_n)
	&= \big(d_E\,\theta^*\big)\big(\rho^{-1}(X_0),\dots,\rho^{-1}(X_n)\big) \\
	&= d\big((\Lambda^n\!\rho^*)(\theta^*)\big)(X_0,\dots,X_n)
}
for sections $X_i\in \Gamma(TM)$. 
The relation \eqref{diffs} describes how exact terms translate in general.
As an example, for the partial derivative ($n=0$) this locally gives
\eq{\label{parD}
	D_\alpha = \rho(\epsilon_\alpha) = \rho^a{}_\alpha\,\partial_a \,.
}
We will come back to this in the next section, 
when we discuss the effect of the field redefinition on  the $H$-flux.


\subsubsection*{Summary}
\label{geom_summary}

We now summarize 
the relevant formulas connecting the
differential geometric quantities of the Lie algebroid $E$ to the standard geometric framework on
the tangent space $TM$:
\eq{
\arraycolsep2pt
\renewcommand{\arraystretch}{1.4}
\begin{array}{@{}l@{\hspace{30pt}}lcl@{}}
{\rm metric} & \fa g_{\alpha\beta}  &=&
\rho^a{}_{\alpha}\, \rho^b{}_{\beta}\, G_{ab}\,, \\
 \mbox{LC connection} &
 	\widehat{\Gamma}^\gamma{}_{\alpha\beta} &=& 
		(\rho^{-1})^{\gamma}{}_c\,\rho^a{}_\alpha\,\rho^b{}_\beta\,\Gamma^c{}_{ab}
		+(\rho^{-1})^\gamma{}_b\,\rho^a{}_\alpha\,\partial_a \rho^b{}_\beta\,, \\
\mbox{curvature tensor} &
 	\widehat{R}^\alpha{}_{\beta\gamma\delta}  &=& 
	  (\rho^{-1})^\alpha{}_a\,\rho^b{}_\beta\,\rho^c{}_\gamma\,\rho^d{}_\delta\,
		R^a{}_{bcd} \,, \\		
\mbox{Ricci tensor} &
	\widehat{R}_{\alpha\beta} &=& \rho^a{}_\alpha\,\rho^b{}_\beta\, R_{ab} \,,\\
\mbox{Ricci scalar} & \widehat{R} &=& R \,, \\
\mbox{partial derivative} &
	D_\alpha &=&  \rho^a{}_\alpha\,\partial_a \,.
\end{array}
}
As one can see, except for the coefficients of the Levi-Civita connection all the expressions
are related simply by applying the anchor map $\rho$.


\subsection{Gauge transformations}
The objects discussed so far behave as tensors under coordinate changes, cf.\ 
section~\ref{sec_alggeom}. However, applying the anchor generically imposes a dependence
on the $B$-field upon the redefined objects.
For this reason and for covering all the symmetries of the string action \eqref{stringaction}, 
we have to study how gauge 
transformations translate under a field redefinition.

We consider the redefinition of the standard Kalb-Ramond field $B$ 
\eq{ \label{gaugetransformationb}
	B=\big(\Lambda^2\rho^*\big)(\mathfrak{b})  \,,
}
with $\mathfrak{b}\in\Gamma(\Lambda^2 E^*)$, which in local coordinates  reads
\eq{\label{frd}
	B_{ab}=(\rho^*)_a^\alpha\,(\rho^*)_b^\beta\;\mathfrak{b}_{\alpha\beta} \,.
}
Note that for our case of interest, namely the field redefinition 
\eqref{GBgb}, the $\mathfrak b$-field takes the form
\eq{
	\mathfrak{b} = \hat{\gamma}\,\hat{\delta}^t-\widehat{G} \,,
}
where the matrices $\hat\gamma$ and $\hat\delta$ generically 
depend on $\widehat{G}$ and $\widehat{B}$.
The
gauge transformations for $B$ read
\eq{	B \to B+d\xi
}
with $\xi$ denoting a
one-form. Let us stress that, since the anchor generically  depends on 
$G$ and $B$, any object containing the
anchor transforms under gauge transformations. Thus, we have  to carefully
distinguish objects whose gauge dependence just stems from the anchor from those
that  are inherently gauge dependent.

Due to the inherent $B$ dependence of $\mathfrak b$ \eqref{frd},  
its overall variation $\delta_\xi \mathfrak b$ under gauge transformations 
receives a contribution according to
\eq{
	\big(\Lambda^2\rho^*\big)(\mathfrak{b}+\widehat{\delta}_\xi \mathfrak{b}) = B+d\xi\,.
}
Inverting \eqref{diffs}, we find
\eq{
	\widehat{\delta}_\xi \mathfrak{b} = \big(\Lambda^2\rho^t\big)(d\xi) = d_E(\rho^t\xi) \,.
}
A second contribution comes from the possible $B$-dependence of
the anchor so that overall we get
\eq{\label{gaugeb}	\delta_\xi
  \mathfrak{b} = \Delta_\xi^2(B) 
  + \widehat{\delta}_\xi \mathfrak{b} \,.
}
Here we introduced the variation of the anchor as
$\Delta_\xi^{n}=\delta_\xi(\otimes^n\rho^t)$.
Since the metric $G$ is gauge invariant, a non-trivial
gauge variation of $g$ can only arise via the anchor so that
\eq{\label{gaugeg}	\delta_\xi g =
  \delta_\xi\big((\otimes^2\rho^t)(G)\big)  = \Delta_\xi^{2}(G) +
 (\otimes^2\rho^t)(\delta_\xi G) = \Delta_\xi^{2}(G) \;.
}

We want to implement the appearance of non-trivial gauge variations
related to the $B$-dependence of the anchor in a consistent modified 
tensor calculus. For this purpose we will introduce the notion of 
\td{tensors}.\footnote{Note that this generalizes the concept of
  $\beta$-tensors introduced for the specific non-geometric frame studied 
 in \cite{Blumenhagen:2012nk,Blumenhagen:2012nt}.} 
In particular, we require the metric to be a \td{tensor}.
This suggests to
define such a tensor by its relation to a gauge invariant object on
$TM$. More precisely, we make the following definition:
\begin{itemize}	

\item[]{\bf Definition:} A section $\tau\in\Gamma\bigl( (\otimes^r E) \otimes
 (\otimes^s E^*) \bigr)$ of the Lie algebroid $E$ is called a \emph{\td{tensor}} if 
\eq{
  \label{gentendef}
    \big[ (\otimes^r\rho)\otimes(\otimes^s\rho^*)\big](\tau)
    \in\Gamma\big( (\otimes^r TM) \otimes (\otimes^s  T^*M) \big)	
}
is gauge invariant.  A \emph{\td{gauge} transformation} of an $n$-form
$\tau\in\Gamma(\Lambda^nE^*)$ is characterized by an $(n-1)$-form $\mathfrak a\in\Gamma(\Lambda^{n-1}E^*)$ as
\eq{
  \label{gengaugetrafo}
  \tau\to \tau + d_E\mathfrak a \,.
}

\end{itemize}
In other words, this definition characterizes a \td{tensor} $\tau$ as an object 
whose image under the anchor map is invariant under $B$-field gauge transformation.
Written in components, a section
$\tau^{\alpha_1\dots\alpha_r}{}_{{\beta_1\dots\beta_s}}$ is a \td{tensor} if
there exists  a gauge invariant standard  $(r,s)$-tensor $T$ with
\eq{
	T^{a_1\dots a_r}{}_{b_1\dots b_s} 
	= \rho^{a_1}{}_{\alpha_1}\dots\rho^{a_r}{}_{\alpha_r} \,
	(\rho^*)_{b_1}{}^{\beta_1}\dots(\rho^*)_{b_s}{}^{\beta_s}\,
	\tau^{\alpha_1\dots\alpha_r}{}_{{\beta_1\dots\beta_s}} \,.
}

Note that the differential geometry we constructed above gives
\td{tensors} right away. Indeed, \eqref{connrel} and the
anchor property written as
\eq{	\widehat{\nabla}_st = \rho^{-1}(\nabla_{\rho(s)}\rho(t)) \;, \hspace{50pt}
[  s,t ]_E = \rho^{-1}([\rho(s),\rho(t)]_L) \,
}
imply that the covariant derivative as well as the Lie algebroid bracket
respect the tensoriality.
Equation \eqref{gaugeb} 
shows that $\mathfrak{b}$ is not a \td{tensor} but receives  a defect
$\widehat\delta_\xi \mathfrak{b}$, 
which is related to its inherent gauge dependence. 

Since the algebroid
differential \eqref{algd} is nilpotent, the natural \td{gauge} invariant object
built from $\mathfrak{b}$ is
\eq{
	\Theta = d_E\mathfrak{b}\quad\in\Gamma(\Lambda^3E^*) \,.
}
Using \eqref{algd}, locally this can be written as\,\footnote{Note that the symmetric part of the connection drops out due to the antisymmetrization.}
\eq{
  \label{r-flux_action}
	\Theta_{\alpha\beta\gamma} = \widehat{\nabla}_{[\alpha} \mathfrak{b}_{\beta\gamma]}\,,
}
where we abbreviated $\widehat{\nabla}_{\epsilon_\alpha}\equiv\widehat{\nabla}_{\alpha}$.
As a consequence, the Bianchi identity
\eq{
 \label{eq:bianchi}
	d_E\Theta = 0 \quad\Longleftrightarrow\quad
	\widehat{\nabla}_{[\alpha}\Theta_{\beta\gamma\delta]} =0
}
is satisfied.
Moreover, using \eqref{diffs} we obtain
\eq{\label{RH}
	\Theta = d_E\big((\Lambda^2\rho^t)B\big) = \big(\Lambda^3\rho^t\big)dB = \big(\Lambda^3\rho^t\big) H \,,
}
i.e.\ $\Theta$ is precisely the redefinition of the $B$-gauge invariant field $H$. 
This also confirms that, unlike $\mathfrak{b}$, $\Theta$ is a
\td{tensor}.


\subsubsection*{Remark}

Motivated by the examples appearing in the literature \cite{Andriot:2012wx,Andriot:2012an}, one might also
want  a transformation relating the two-form $B$ to a two-vector
$\beta\in\Gamma(\Lambda^2E)$. This is
different from  \eqref{frd} where $\mathfrak{b}\in\Gamma(\Lambda^2E^*)$, and 
requires a map $\sigma:E\to T^*M$.  This is apparently not the anchor, but recalling \eqref{diagram}, we
do have two natural candidates for such a map:
\eq{
	\sigma_1:E\to T^*M &;\,s\mapsto \rho^*\circ g^\sharp(s) \\
	\sigma_2:E\to T^*M &;\,s\mapsto G^\sharp\circ \rho(s) \,.
}
Using the translation of the
metrics \eqref{genGg} yields
$\sigma_1 = \sigma_2 \equiv \sigma$. Then, the redefinition reads
\eq{	B = (\Lambda^2\sigma)\beta = \big(\Lambda^2(\rho^*\circ
  g^\sharp)\big)\beta 	= \big(\Lambda^2\rho^*\big)\big((\Lambda^2g^\sharp)\beta\big) \,.
}
Hence also this case fits into the general picture \eqref{frd} by
identifying
$\mathfrak{b}=(\Lambda^2g^\sharp)\beta$. This was already used in \eqref{lmubetaB}.


\subsection{The general redefined action}
\label{thegenredef}

In the previous sections, we discussed all relevant ingredients for giving the general action arising from the NS-NS Lagrangian \eqref{stringaction} by redefining the metric and the $B$-field according to \eqref{genGg} and \eqref{frd}.
For the new action, we have to give the new Ricci scalar, the flux term and the dilaton term. Moreover, also the measure changes. Let us start by discussing the latter.

The standard measure behaves under a field redefinition \eqref{genGg} as follows:
\eq{
  \label{measure_07}
	\sqrt{\mdet{G}} 
	= \sqrt{\mdet{((\rho^*)_a{}^\alpha(\rho^*)_b{}^\beta g_{\alpha\beta})}}
	= \sqrt{\mdet{g}}\,\mdet{\rho^*}\,.
} 
The measure is well-defined  by recalling that the anchor is
invertible. 

The remaining terms in the action have been discussed in
\eqref{curvrelloc} for the curvature and in \eqref{RH} for the flux
term. Hence the translation of the Ricci scalar is straightforward. 
For the $H$-flux term we observe that by \eqref{genGg} and \eqref{RH}
\eq{
	H_{abc}\, G^{am}G^{bn}G^{ck}\,H_{mnk} 
	= \Theta_{\alpha\beta\gamma}\, 
		g^{\alpha\mu}g^{\beta\nu}g^{\gamma\rho}\,\Theta_{\mu\nu\rho} \;,
}
where $\Theta$ has been defined in \eqref{r-flux_action}.
Using \eqref{parD}, the dilaton term translates  analogously
\eq{
	\partial_a\phi\, G^{ab}\,\partial_b\phi = D_\alpha\phi\, g^{\alpha\beta} D_\beta\phi \,,
}
where  the dilaton itself does not transform. Note that
\td{scalars}
are related to usual scalars without any contraction with the anchor. As every term in the Lagrangian is a scalar,
each individual term maps to the corresponding \td{scalar} directly. Putting all these
pieces together, we obtain  from the NS-NS action \eqref{stringaction}
the final  action in a non-geometric frame
\eq{\label{redefaction}
	\mathcal{S} = -\frac{1}{2\kappa^2}\int d^nx\sqrt{\mdet{g}}\,\mdet{\rho^*}\,e^{-2\phi}
	\Big(\widehat{R}-\tfrac{1}{12}\,\Theta_{\alpha\beta\gamma}\Theta^{\alpha\beta\gamma}
	+ 4D_\alpha\phi\,D^\alpha\phi\Big) \,.
}
By construction, this action is invariant under diffeomorphisms and 
two-form gauge transformations, whose inherent part acts like a 
$\mathfrak{b}$-gauge transformation
\eq{
	\mathfrak{b}\to \mathfrak{b} + d_E\mathfrak{a}
}
for $\mathfrak{a}\in\Gamma(E^*)$. Hence \eqref{redefaction} bears the
redefined analogs of the symmetries of the geometric action 
\eqref{stringaction}
and provides  the generalization of the action 
\eqref{finalaction} to any non-geometric frame.

Let us emphasize that by construction \eqref{redefaction} and \eqref{stringaction} are directly related by the field redefinition \eqref{genGg} and \eqref{frd}:
\eq{
	\mathcal{S}(g,\mathfrak{b}) \,
	\xleftrightarrow[\mathfrak{b}=(\wedge^2\!\rho^t)
          B]{\,\,g=(\otimes^2\!\rho^t)G\,\,} \,
	S(G,B) \,.
}



\subsubsection*{String action in Frame I}

Let us recall that the anchor in frame I \eqref{fieldrefineA} was given 
 in eq.\ \eqref{anchorI} as
\eq{
	\rho_\mathrm{I} = \mathds{1}-G^{-1}B = \mathds 1 + \widehat \beta \hspace{1pt}\widehat G \,,
}
where for the last step we employed the field redefinitions \eqref{lmubetaB} and \eqref{new_fields_hvm}.
The partial derivative \eqref{parD} of the Lie algebroid then becomes
\eq{
	D_a = \partial_a -  \widehat G_{a\beta}\hspace{1pt}\widehat\beta^{\beta c}\,\partial_c \,.
}
For the measure of the redefined action, we can use the relation derived in equation \eqref{measure_07}, which leads to
\eq{
  \sqrt{\bigl|\widehat G^{-1}\bigr|} \, \bigl|\widehat G^{-1} - \widehat \beta\bigr|^{-1} \,.
}
The components of the flux $\Theta$ can be determined for instance from equation \eqref{RH} 
by recalling that in our conventions $H_{abc} = 3\hspace{1pt} \partial_{[a} B_{bc]}$. We then compute
\eq{
  \label{frameI_18}
  \Theta_{\alpha\beta\gamma} = 3 \hspace{1pt}\bigl( \mathds 1 - \tilde G \hspace{1pt}
   \widehat \beta\bigr)_{[\alpha}{}^{a}\hspace{1pt}\bigl( \mathds 1 - \widehat G \hspace{1pt}
   \widehat \beta\bigr)_{\beta}{}^{b}\hspace{1pt}\bigl( \mathds 1 - \widehat G \hspace{1pt}
   \widehat \beta\bigr)_{\gamma]}{}^{c}
   \;\partial_a \bigl( \widehat G_{b m} \widehat \beta^{mn} \widehat G_{nc} \bigr) \;.
}
As one can see, the non-geometric analog of the $H$-flux is a rather complicated expression. 
However, the flux \eqref{frameI_18} does contain the familiar $R$-flux term 
$R^{abc} =3\hspace{1pt} \widehat\beta^{[a|m}\partial_m \widehat\beta^{bc]}$, which is 
accompanied by a plenitude of additional terms
\eq{
   \Theta_{\alpha\beta\gamma} =-3 \hspace{1pt} \widehat G_{\alpha a}\,  \widehat G_{\beta b} \,
   \widehat G_{\gamma c} \,
   \bigl[ \widehat \beta^{[a|m}\partial_m \widehat \beta^{|bc]} \bigr]
   + \mathcal O(\partial \widehat G) + \mathcal O ( \partial \widehat\beta)\;.
}
When expressing the Ricci scalar in terms of the fields $\widehat G$ and $\widehat\beta$, we obtain similarly 
involved expressions, and we refrain from presenting them here. 
The explicit form of the action in the $(\widehat G,\widehat \beta)$-frame, modulo total-derivative terms, 
can be found in \cite{Andriot:2011uh,Andriot:2012wx,Andriot:2012an}.


\subsubsection*{String action in Frame II}

For our second example we recall that the field redefinition was given in \eqref{frameBredefine}.
Furthermore, the corresponding anchor $\tilde\rho_\mathrm{II}$ for a Lie algebroid on $T^*M$ has
been derived in  \eqref{anchorII} 
\eq{
	\tilde{\rho}_\mathrm{II}= - \hat{\beta} \,,
}
where $\hat\beta$ is an antisymmetric bi-vector.
The associated partial derivative can  be determined as
\eq{
  D^{\alpha} = \hat\beta^{\alpha b}\partial_b \,.
}
The measure for the action in the redefined field variables can be inferred for instance from 
\eqref{frameBredefine} and takes the form
\eq{
  \sqrt{\mdet{\hat g}} \; \bigl| \hat \beta  \bigr|^{-1} \,.
}
For the flux $\Theta$ we employ again the relation shown in \eqref{RH} which, using \eqref{frameBredefine}, allows us to write
\eq{
  \Theta^{\alpha\beta\gamma} = -3 \hspace{1pt} \hat\beta^{[\alpha|m}\partial_m \hat\beta^{|\beta\gamma]}
  \;.
}
The curvature scalar can be constructed along the lines outlined above, as was done in \cite{Blumenhagen:2012nk,Blumenhagen:2012nt}. 
Using these building blocks in \eqref{redefaction}, one can construct the 
action \eqref{finalaction} in the non-geometric
$(\hat g, \hat\beta)$-frame \cite{Blumenhagen:2012nk,Blumenhagen:2012nt}.


\section{Further aspects of non-geometric gravity}

In this section we discuss a couple of interesting aspects of
the generalized gravity action \eqref{redefaction} in the non-geometric frame.
First, we will discuss how it fits into the formalism
of double field theory (DFT), which is a candidate to provide
a unified framework for the geometric and non-geometric phases 
 of (bosonic) string theory, at least at tree-level. 
Second, we  apply the formalism developed above  to perform the
translation of the remaining terms in the superstring action. This 
includes terms from the Ramond-Ramond sector as well as 
fermionic terms. We also comment
on higher order $\alpha'$-corrections and  the tree-level equations of
motion of the action \eqref{redefaction}. Finally, we discuss
the important question, in which sense non-geometric frames are
appropriate or  useful
to describe non-geometric backgrounds.

\subsection{Relation to double field theory}
\label{sec_dft}

The goal of this section is  whether and how the 
action in the non-geometric frames
\eqref{redefaction} does   arise in DFT. 
In the following, we describe how this works
for the case of rigid $O(D,D)$ transformations.
Non-geometric frames related to $O(D,D)$-transformations which contain generic
local $\beta$-transformations go beyond the regime of the DFT action and involve
field redefinitions which cannot be generated by its
symmetries.

\subsubsection*{Basics of DFT}

In DFT not only the dimension of the bundle is doubled
but even the number of coordinates.
This is done by also introducing the canonical conjugate
variables for the string winding operators, which are called
winding coordinates $\tilde x_i$, and arranging
them into a doubled vector $X^M=(\tilde x_i,x^i)$.

As was developed in \cite{Hohm:2010pp,Hull:2009mi,Hull:2009zb,Hohm:2010jy},
one can formulate an action on this doubled space in which the
generalized metric appears explicitly\,\footnote{Note that usually, one splits the $n$-dimensional space-time into a $D$-dimensional compact part, and an $(n-D)$-dimensional non-compact part. The doubling of coordinates 
takes place only in the compact space, the action for the other coordinates is unchanged. This
 is implicitly assumed in \eqref{dftaction}.}
\begin{eqnarray}
 \label{dftaction}
  && S_{{\rm DFT}}=-{1\over 2\kappa^2}
\int 
d^Dx\, d^D\tilde x \; e^{-2d}\, \biggl( \tfrac{1}{ 8} \hspace{1pt}{\cal H}^{MN} (\partial_M
     {\cal H}^{KL}) (\partial_N {\cal H}_{KL}) \\
   &&\hspace{20pt}  -\tfrac{1}{ 2} \hspace{1pt}{\cal H}^{MN} (\partial_N
     {\cal H}^{KL})( \partial_L {\cal H}_{MK}) 
     -2 \hspace{1pt}(\partial_M d) (\partial_N {\cal H}^{MN})
   + 4\hspace{1pt} {\cal H}^{MN}  (\partial_M d)  (\partial_N d) \biggr) .
  \nonumber
\end{eqnarray}
Note that here $\partial_M=(\tilde\partial^i,\partial_i)$, and $d$ denotes the dilaton which is defined as 
$\exp(-2d)=\sqrt{\mdet{G}} \exp(-2\phi)$.
This action has been determined by invoking a number of symmetries:
First it was required to be invariant under local diffeomorphisms of the
coordinates $X^M$, i.e.\ $(\tilde x_i,x^i)\to (\tilde
x_i+\tilde\xi_i(X),x^i+\xi^i(X))$\, \footnote{
The $x^i$ dependence of these two diffeomorphisms include  both 
standard diffeomorphisms and $B$-field gauge transformations. Note that the 
winding coordinate dependence of $\xi^i$ also gives what one might call
$\beta$-field gauge transformations.}. Second, the action is invariant
under a global or rigid $O(D,D)$ symmetry, which acts as
\eq{
\label{dft_trafo}
         {\cal H}'&=h^t\, {\cal H} h \,, \qquad  d'=d\;,\\
         X'&=h X\,, \qquad \partial'=(h^t)^{-1} \,\partial \;,
}
with \footnote{The constant matrix $\mathfrak b$ should not be confused with
  the space-time dependent field introduced in \eqref{gaugetransformationb}.}  
\eq{      
            h=\left(\begin{matrix} \mathfrak a & \mathfrak b\\ 
                             \mathfrak c & \mathfrak d \end{matrix}\right).
}
For  this manifest $O(D,D)$ invariance this action has to be
supplemented by the so-called strong constraint 
\eq{
  \label{strong_c}
      \partial_i A\, \tilde\partial^i B +  \tilde\partial^i A\, \partial_i B=0\;, 
}
with $A$, $B$ arbitrary fields. Whether this constraint can be weakened
in compactifications of DFT has recently been analyzed in \cite{Geissbuhler:2013uka}.
Solving \eqref{strong_c} by setting to zero the
derivative with respect to the winding coordinates $\tilde\partial^i=0$, the
double field theory 
action reduces to  the action in the geometric 
frame \eqref{stringaction}.

Let us also recall an alternative formulation of DFT.
One  introduces the $O(D,D)$ covariant partial derivatives
\eq{ \label{curlyD}
{\cal D}_i =&\, \partial_i - {\cal E}_{ik}\tilde \partial^k \,,\\
\overline {\cal D}_i =&\, \partial_i + {\cal E}_{ki} \tilde \partial^k \;,
}
with the background matrix ${\cal E}$ defined as  
\eq{
{\cal E}_{ij} =\, G_{ij} + B_{ij} \,.
}    
The DFT action  \eqref{dftaction} can be expressed  as \cite{Hohm:2010jy} 
\eq{
\label{dft}
S_{\mathrm{DFT}} \hspace{-0.7pt}&=  \int \hspace{-0.2pt}d^Dx\, d^D\tilde x\; {\cal L}_{\mathrm{DFT}}({\cal E},{\cal D},d) \\
&=  \int \hspace{-0.5pt}d^Dx\,d^D\tilde x \, e^{-2d} \Big[
  \!-{\textstyle\frac{1}{4}} G^{ik}G^{jl}G^{pq} \bigl( {\cal D}_p {\cal
    E}_{kl}\, {\cal D}_q {\cal E}_{ij} - {\cal D}_i {\cal E}_{lp}\, {\cal D}_j
     {\cal E}_{kq}  - \overline{\cal D}_i {\cal E}_{pl}\, \overline{\cal D}_j
     {\cal E}_{qk}\bigr) \\
&\hspace{85.6pt}
+ G^{ik}G^{jl} \left({\cal D}_i d\, \ov{\cal D}_j {\cal E}_{kl} + \ov{\cal D}_i d\, {\cal D}_j {\cal E}_{lk} \right) 
+ 4 G^{ij}{\cal D}_i d\, {\cal D}_j d \Big] .
}
The rigid $O(D,D)$ symmetry $X'=h\, X$ acts as follows
\eq{     {\cal E}'&=(\mathfrak a{\cal E} +\mathfrak b)(\mathfrak c+{\cal E}\mathfrak d)^{-1}\, ,\\
         {\cal D}_i &= M_i{}^j\, {\cal D}'_j \;, \qquad \ov{\cal D}_i =\, \ov
         M_i{}^j\, \ov{\cal D}'_j\;,
}
where the matrices $M,\ov M$ are given by 
\eq{
M =(\mathfrak d - \mathfrak c{\cal E}^t)^t  \;, \quad \ov M = (\mathfrak d + \mathfrak c {\cal E} )^t \;.
}
As it will become relevant soon, we  also provide the implied
transformation of the metrics
\eq{
\label{hannover96}
            G=\ov M\, G'\, \ov M^t\, .
}

The idea now is that the actions in the non-geometric frames
correspond to different solutions of the strong constraint. 
The latter allows us to express the winding derivative in terms of
the usual derivative. However, implementing this constraint and
directly reducing the DFT action
is not a trivial task so that we use the rigid $O(D,D)$ symmetry
to rotate the solution of the strong constraint again to the simple
form $\tilde\partial=0$ and then perform the reduction.


\subsubsection*{Relation of DFT to non-geometric actions}

To connect to our analysis from previous sections, 
our starting point is DFT with the fields 
$\widehat{\cal E} = \widehat G + \widehat B$ and
an action ${\cal L}_{\rm DFT}(\widehat{\cal E},\widehat{\cal D},\hat d)$.
Now we are solving the strong constraint by an ansatz which contains the matrices used for the field redefinition of section \ref{sec_oddredef} in the special case of constant $a,b$:   
\eq{ \label{sol_sc}
\hat{\tilde\partial}^i = (b^t)^{ij}\, \partial_j\, , \qquad \hat\partial_i = (a^t)_i{}^j\, \partial_j \,.
}
Indeed the strong constraint \eqref{strong_c} becomes
\eq{
	\partial_iA\,\partial_jB\,(b\,a^t+a\, b^t)^{ij} =0 \;.
}
Instead of reducing the DFT Lagrangian to a Lagrangian ${\cal L}(\widehat{\cal
  E},\partial,\hat d)$ depending on only  half of the coordinates we use the rigid $O(D,D)$ symmetry to transform it to a frame
where the strong constraint is simply solved by $\tilde\partial^i = 0$. It will turn out below that the right choice for the $O(D,D)$ transformation
is
\eq{\label{rightchoice}
          h=({\cal M}^t)^{-1}=\left(\begin{matrix} d & c\\ 
                             b & a \end{matrix}\right)\;,
}
where ${\cal M}$ is the $O(D,D)$ matrix we used for the field
redefinitions and the definition of the anchor in the previous sections.
The fields in the new frame are denoted as 
${\cal E} = G +  B$.
Using \eqref{dft_trafo} we find that the partial derivatives transform as follows
\eq{ 
\label{rotatedpartials}
\begin{pmatrix}
\tilde{\partial} \\
\partial 
\end{pmatrix} = {\cal M} 
\begin{pmatrix}
\hat{\tilde\partial} \\
\hat\partial
\end{pmatrix} =\, \begin{pmatrix}
a & b \\
c & d 
\end{pmatrix} 
\begin{pmatrix}
\hat{\tilde\partial} \\
\hat\partial
\end{pmatrix}\,.
}
Therefore, the solution \eqref{sol_sc} to the strong constraint
simply becomes $\tilde\partial^i=0$ in the new coordinates. 
Moreover, from \eqref{hannover96} we get the relation between the old and
the new metric
\eq{
            \widehat G=\big(a^t+(\widehat G-\widehat B)\, b^t\big)\, G\,
                       \big(a^t+(\widehat G-\widehat B)\, b^t\big)^t \;,
}
which precisely agrees with the equations
\eqref{GBgb} and \eqref{invgamma} for the field redefinition
relating the geometric frame to the non-geometric one. 
The same can be shown for the two-form. This justifies the choice of the $O(D,D)$-transformation \eqref{rightchoice}.
Since the dilaton $d$ is invariant, the measure factor is
\eq{
e^{-2d} =\, \sqrt{\mdet{G}} e^{-2\phi}=
\sqrt{\bigl|\widehat G\bigr|}\mdet{\hat\gamma^{-1}} e^{-2\phi} \;.
}
Therefore, we can conclude that reducing DFT in the new frame with  
$\tilde \partial^i  =0$ results in the
standard NS-NS action  $S_{\mbox{\scriptsize NS-NS}}(G,B,\partial_i ,\phi)$ 
with redefined background fields 
$G(\widehat G,\widehat B)$ and $B(\widehat G, \widehat B)$.
But as was shown in section \ref{thegenredef}, this action is equivalent 
to the string action \eqref{redefaction} in the non-geometric frame.


\subsection{Relation to supergravity}
After having considered the NS-NS sector to lowest
order in $\alpha'$, let us now turn to the remaining terms
in the low-energy effective action of string theory.
In the following we want to discuss how the constructions given above apply to Ramond-Ramond (R-R) and fermionic fields.
It turns out that we can translate the whole supergravity action to the non-geometric frames.


\subsubsection*{General remarks}

Let us recapitulate the necessary ingredients. In section~\ref{sec_ngdiffgeo}
we have seen that any tensor in the standard frame becomes a \td{tensor}
in the redefined theory if the anchor is applied to all the indices,
cf.\ \eqref{gentendef}. Since the $O(D,D)$ induced field redefinition only
concerns the metric and the $B$-field, any tensor invariant under $B$-gauge
transformations becomes a \td{tensor} by anchoring. This applies in
particular to fields which transform under gauge transformations different
from $B$-gauge transformations, e.g.\ Ramond-Ramond fields. Hence we transform
every $(r,s)$-tensor $T$ on $TM$ to an 
$(r,s)$-\td{tensor} $\widehat{T}$ on the Lie algebroid $E$ via
\eq{\label{tentran1}
	\widehat{T}^{\alpha_1\dots\alpha_r}{}_{\beta_1\dots\beta_s} =
	(\rho^{-1})^{\alpha_1}{}_{a_1}\dots(\rho^{-1})^{\alpha_r}{}_{a_r}\,
	(\rho^t)_{\beta_1}{}^{b_1}\dots(\rho^t)_{\beta_s}{}^{b_s}\,
	T^{a_1\dots a_r}{}_{b_1\dots b_s} \,.
}
Using \eqref{connrel} this also holds for covariant derivatives
\eq{\label{tentran2}
	\widehat{\nabla}_\gamma \widehat{T}^{\alpha_1\dots\alpha_r}{}_{\beta_1\dots\beta_s} =
	(\rho^t)_\gamma{}^c\,&(\rho^{-1})^{\alpha_1}{}_{a_1}\dots(\rho^{-1})^{\alpha_r}{}_{a_r}\times\\
	&(\rho^t)_{\beta_1}{}^{b_1}\dots(\rho^t)_{\beta_s}{}^{b_s}\,
	\nabla_c T^{a_1\dots a_r}{}_{b_1\dots b_s} \,,
}
which is just a special case of \eqref{tentran1}.
Note that all terms appearing in the Lagrangian as well as in the equations of
motion are tensors which do not transform under $B$-gauge
transformations. Thus \eqref{tentran1} and \eqref{tentran2} suffice to
translate every term. In addition,  the measure transforms according to
\eqref{measure_07}, so the appropriate determinant appears in the
action. The following is the direct generalization of the results of
\cite{Blumenhagen:2012nt}.

\subsubsection*{R-R sector}
The Ramond-Ramond fields in e.g.\ type IIA supergravity are the
antisymmetric tensors $C_1$ and $C_3$, with corresponding field
strengths\footnote{\label{foot:rrcov}Note that they are usually defined
with the partial instead of the covariant derivative, but the usual
symmetric connection coefficients drop out in the antisymmetrization.
Our `hatted' Christoffel symbols however are in general not symmetric.}
\eq{
	F_{a_1a_2} &= 2\nabla_{[a_1}C_{a_2]} \,,\\
	F_{a_1a_2a_3a_4} &=4 (\nabla_{[a_1}C_{a_2a_3a_4]}+C_{[a_1}H_{a_2a_3a_4]}) \,.
}
The Lagrangian for these fields is given by:
\eq{
	\mathcal{L}^{\text{R-R}}_{\mathrm{IIA}} \sim
	&\frac14F_{a_1a_2}F^{a_1a_2}+\frac{1}{48}F_{a_1a_2a_3a_4}F^{a_1a_2a_3a_4} \\
	&-\frac{i}{144\cdot\sqrt{\mdet{G}}}\epsilon^{a_1\ldots a_{10}}
		\nabla_{[a_1}C_{a_2a_3a_4]}\nabla_{[a_5}C_{a_6a_7a_8]}B_{a_9a_{10}} \,,
}
where $\epsilon^{a_1\ldots a_{10}}$ is the antisymmetric symbol (which
is not a tensor).
By applying \eqref{tentran1}, we can express this Lagrangian in the redefined fields on the Lie algebroid 
\eq{
	\mathcal{L}^{\text{R-R}}_{\mathrm{IIA}}=\hat{\mathcal{L}}^{\text{R-R}}_{\mathrm{IIA}} \sim
	&\frac{1}{4}\widehat F_{\alpha_1\alpha_2}\widehat F^{\alpha_1\alpha_2}
		+\frac{1}{48}\widehat F_{\alpha_1\alpha_2\alpha_3\alpha_4}
		\widehat F^{\alpha_1\alpha_2\alpha_3\alpha_4}\\
	&-\frac{i}{144\cdot\sqrt{\mdet{g}}}\epsilon^{\alpha_1\ldots\alpha_{10}}
		\widehat\nabla_{[\alpha_1}\widehat C_{\alpha_2\alpha_3\alpha_4]}
		\widehat\nabla_{[\alpha_5}\widehat C_{\alpha_6\alpha_7\alpha_8]}
		\mathfrak b_{\alpha_9\alpha_{10}}\,,
}
where the corresponding action includes the measure \eqref{measure_07}.

These redefined R-R fields are invariant under the usual gauge transformations
\eq{
&\delta_{\Lambda_{(0)}}\widehat C_{\alpha_1}=\widehat\nabla_{\alpha_1}\Lambda,\qquad
\delta_{\Lambda_{(0)}}\widehat C_{\alpha_1\alpha_2\alpha_3}=-\Lambda\Theta_{\alpha_1\alpha_2\alpha_3},\\
&\delta_{\Lambda_{(2)}}\widehat{C}_{\alpha_1\alpha_2\alpha_3}=\widehat{\nabla}_{[\alpha_1}\Lambda_{\alpha_2\alpha_3]},
}
where $\Lambda_{(0)}$ and $\Lambda_{(2)}$ are arbitrary zero- and two-forms. The invariance under $\delta_{\Lambda_{(0)}}$ follows from the Bianchi identity \eqref{eq:bianchi}.

\subsubsection*{Fermionic sector}
In the following, Greek indices starting with $\mu$ will denote Lorentz indices, whereas Greek indices starting with $\alpha$ are Lie algebroid indices and Latin indices are $TM$ indices, as before.
To write down the Lie algebroid action for R-NS and NS-R fields we need to consider vielbein fields $\hat e_\alpha^\mu$ which fulfill a relation analogous to the normal frame fields $e_a^\mu
$:
\eq{
	\hat e^\mu_\alpha\, \hat e^\nu_\beta\, g^{\alpha \beta}
	= \delta^{\mu \nu}=e^\mu_a\, e^\nu_b\, G^{ab}.
}
By \eqref{gGloc}, this implies $\hat e^\mu_\alpha=\rho^a{}_\alpha e^\mu_a$.

With these vielbeins, we can build a spin connection
$\widehat\omega$ on $E$ using the standard formula:
\eq{
	\widehat{\omega}_\gamma{}^\mu{}_\nu
	=\hat e_\alpha^\mu\,\hat e^\beta_\nu\,\hat\Gamma^\alpha{}_{\beta\gamma}
		+\hat e^\mu_\alpha\, D_\gamma \hat e^{\alpha}_{\nu} \,.
}
Using \eqref{Christoffel}, we can write this as
\eq{
	\widehat{\omega}_\gamma{}^\mu{}_\nu=\rho^c{}_\gamma\, \omega_c{}^\mu{}_\nu \,,
}
where
\eq{\label{eq:vielbeins}
	\omega_c{}^\mu{}_\nu=e_a^\mu\, e^b_\nu\,\Gamma^a{}_{bc}+e^\mu_a\,\partial_c e^a_\nu
}
is the standard spin connection on $TM$.

Now consider the fermionic sector of type IIA supergravity (which can be found in \cite{Giani:1984wc}). It contains the dilaton $\phi$, the gravitino $\psi_a$, the dilatino $\lambda$, R-R fields $F_{a_1,\ldots a_k}$, covariant derivatives $D_a$ and gamma matrices $\gamma_a$. The gamma matrices on $TM$ are given by $\gamma_a=e^\mu_a \gamma_\mu$, so we can consistently define
\eq{
	\hat\gamma_\alpha=\hat e^\mu_\alpha\, \gamma_\mu
	=\rho^a{}_\alpha\, e^\mu_a\, \gamma_\mu=\rho^a{}_\alpha\, \gamma_a \,.
}
The kinetic term of the dilatino has the following form:
\eq{
	\mathcal{L}^\lambda_{\mathrm{IIA}}\sim\bar\lambda\gamma^a
	\left(\partial_a-\frac{i}{4}\omega_a{}^{\mu\nu}\gamma_{\mu\nu}\right)\lambda \,,
}
where $\gamma_{\mu_1\ldots\mu_k}=\gamma_{[\mu_1}\ldots\gamma_{\mu_k]}$. Because $\lambda$ is a spinor, the spin connection $\omega$ has to be included. Since the dilatino does not have vector indices, we have $\hat\lambda=\lambda$, so we can write
\eq{
	\hat{\mathcal{L}}^{\hat\lambda}_{\mathrm{IIA}}\sim
	\bar{\hat\lambda}\hat\gamma^\alpha 
	\left(D_\alpha-\frac{i}{4}\widehat{\omega}_\alpha{}^{\mu\nu}\gamma_{\mu\nu}\right) \hat\lambda \,.
}
The kinetic term of the gravitino looks like
\eq{
	\mathcal{L}^\psi_{\mathrm{IIA}}\sim\bar\psi_a\gamma^{abc}
	\left(\nabla_b-\frac{i}{4}\omega_b{}^{\mu\nu}\gamma_{\mu\nu}\right)\psi_c
}
because $\psi_a$ has a form index in addition to the (implicit) spinor
indices.\footnote{Note that the Christoffel symbols drop out in the
  Lagrangian due to their symmetry. In the new frame, they are (in general)
  not symmetric, so we have to keep them (cf.\ footnote \ref{foot:rrcov}).}
Here \eqref{tentran1} reads $\hat\psi_\alpha=(\rho^t)_\alpha{}^a\psi_a$, and with the above the transformed Lagrangian is
\eq{
	\hat{\mathcal{L}}^{\hat\psi}_{\mathrm{IIA}}\sim
	\bar{\hat{\psi}}_\alpha\gamma^{\alpha\beta\gamma}
	\left(\widehat{\nabla}_\beta-\frac{i}{4}\widehat{\omega}_\beta{}^{\mu\nu}\gamma_{\mu\nu}\right)
	\hat\psi_\gamma \,.
}
Again, the equivalence of the gravitino actions in both frames follows from \eqref{gentendef}.


\subsection{Higher order corrections}

The action \eqref{stringaction} is the lowest order contribution in the string
tension $\alpha'$ to the effective action of the massless modes $G$, $B$ and
$\phi$. Although the higher order corrections are not unique due to a freedom
of redefining the fields, all terms can be composed of (covariant derivatives
of) the curvature tensor $R^a{}_{bcd}$, the three-form $H$, the dilaton
$\partial_a\phi$ and contractions thereof  \cite{Metsaev:1987bc,Metsaev:1987zx,Hull:1987yi}.

For the translation of these higher order corrections to a non-geometric frame, we note that all terms 
in the action are scalars which are invariant under gauge transformations of the Kalb-Ramond field.
To obtain \td{scalars} on the Lie algebroid, we therefore just have to perform the replacements
\eq{
	R^a{}_{bcd} \quad&\to\quad \widehat{R}^{\,\alpha}{}_{\beta\gamma\delta} \,,\\
	H_{abc} \quad&\to\quad \Theta_{\alpha\beta\gamma} \,,\\
	\partial_a\phi \quad&\to\quad D_\alpha\phi \,,\\
	\sqrt{\mdet{G}} \quad&\to\quad \sqrt{\mdet{g}}\mdet{\rho^*} \,,
}
cf.\ \eqref{curvrelloc}, \eqref{RH} and \eqref{measure_07}. Indeed, contractions
of the latter fields are then \td{scalars}. The resulting terms contribute
as higher order $\alpha'$-corrections to the action \eqref{redefaction} in
the NS-NS sector, and are related by the
general field redefinition \eqref{genGg} and \eqref{frd} to the actions in the
usual frame.


\subsection{Equations of motion}

The recipe applied above is also suitable for the equations of motion of the
action \eqref{redefaction}. The explicit computation is very cumbersome, but
we can equally well just transform the well-known equations of motion for
\eqref{stringaction}. Again, every term therein is a gauge invariant tensor and anchoring it
gives \td{tensors}. As the anchor is a bijection, we can just drop the
overall anchor factors which yields an independent set of equations for
$\widehat{G}$, $\mathfrak{b}$ and $\phi$. This way we obtain the equations
of motion for the general redefined action \eqref{redefaction}
\eq{\label{eom_gen}
	0 &= \widehat{R}_{\alpha\beta} + 2\,\widehat{\nabla}_\alpha\widehat{\nabla}_\beta\phi
			-\frac{1}{4}\,\Theta_{\alpha\mu\nu}\, \Theta_\beta{}^{\mu\nu} \,,\\
	0 &= -\frac{1}{2}\,g^{\alpha\beta}\,\widehat\nabla_\alpha\widehat{\nabla}_\beta\phi 
			+g^{\alpha\beta}\,\widehat\nabla_\alpha\phi\widehat{\nabla}_\beta\phi 
			-\frac{1}{24}\,\Theta_{\alpha\beta\gamma}\, \Theta^{\alpha\beta\gamma} \,,\\
	0 &= \frac{1}{2}\,\widehat{\nabla}^\mu\Theta_{\mu\alpha\beta} 
			-(\widehat{\nabla}^\mu\phi)\Theta_{\mu\alpha\beta} \,. 
}
Let us emphasize  that \eqref{eom_gen} are the equations of motion for the
action \eqref{redefaction} in
an arbitrary non-geometric frame. Here, we considered $\mathfrak{b}$ instead of $\widehat{B}$ for simplicity; the appearance of the
former in \eqref{redefaction} is analogous to $B$ in \eqref{stringaction}.


\subsection{Non-geometric frames  --  non-geometric backgrounds}

The notion of non-geometry applies to string theory backgrounds which elude a
description in terms of usual manifolds. In ordinary geometry, the transition
functions between local patches of a manifold are diffeomorphisms, possibly
accompanied by gauge transformations. These are encoded in the geometric
group $G_\mathrm{geom} = G_{d \Lambda}\rtimes G_\mathrm{diffeo}$, the local
symmetry group for the string action \eqref{stringaction}. For patching up
non-geometric backgrounds, however, a transformation beyond $G_\mathrm{geom}$ is
necessary. Hence, for identifying a non-geometric background global properties
have to be taken into account. Concrete examples of non-geometric backgrounds
arise from  T-dualizing geometric ones. The 
\emph{T-fold} introduced in \cite{Hull:2004in} gives such an example for which
the structure group contains general $O(D,D;\mathbb{Z})$ transformations.

In the previous sections, we have given a description
of string theory in  general non-geometric \emph{frames}.
Here, different frames were defined by applying $O(D,D)$ transformations
to a given generalized metric. The question now arises whether
and how the description of a \emph{given} non-geometric 
background might simplify
by choosing an appropriate non-geometric frame.
As one knows from the standard $Q$-flux background, the concrete expressions
for the backgrounds fields might simplify, but the essential question
is whether the transition functions can become members of
the symmetry group in a non-geometric frame.

To analyze this question, let us consider the generalized metric \eqref{genmetric}. Suppose 
$\mathcal{H}_1$ and $\mathcal{H}_2$ are the generalized metrics in two
overlapping patches of a 
non-geometric background with the transition function  given  
by $\mathcal{T}\notin G_\mathrm{geom}$
\eq{
	\mathcal{H}_1 = \mathcal{T}^t\mathcal{H}_2\mathcal{T} \,.
}
Now, going to another frame by applying an $O(D,D)$ transformation 
$\mathcal{M}$ to this background, the transition 
function $\mathcal{T}$ changes to
\eq{\label{trantran}
	\mathcal{T}' = \mathcal{M}^{-1}\mathcal{T}\mathcal{M} \,.
}
However, performing a field redefinition based on $\mathcal{M}$ also changes
the geometric group  which, as we have seen,  is the symmetry group of the
action \eqref{redefaction} in this non-geometric frame.
The new symmetry group becomes
$G_\mathrm{geom}'=\mathcal{M}^{-1}G_\mathrm{geom}\mathcal{M}$ so that 
\eq{
	\mathcal{T}'\notin G_\mathrm{geom}' \quad\Longleftrightarrow\quad 
	\mathcal{T}\notin G_\mathrm{geom} \,,
}
i.e.\ the transition function remains to be non-geometric.

\subsubsection*{$Q$-flux example}

As an example, we consider the approximate
$Q$-flux background \cite{Shelton:2005cf}. It arises from a three-torus
parametrized by coordinates $(x,y,z)$ with constant $H$-flux $N$ by performing
two T-dualities in the isometric directions, say $x$ and $y$. The background is given by
\eq{\label{qback}
	G = \frac{1}{1+N^2z^2}\big(dx^2+dy^2\big)+dz^2 \, ,\quad
	B = \frac{Nz}{1+N^2z^2}\, dx\wedge dy \,,
}
where we have set the radii of the torus to one. The $z$-direction is a cycle of the
torus and as such admits a periodicity
$z\mapsto z+k$ for $k\in2\pi\mathbb{Z}$. However, the fields \eqref{qback} are not periodic 
and the change in $G$ and $B$ cannot be compensated by a diffeomorphism or a gauge
transformation. Instead, the required transformation is given by a $\beta$-transform
\eq{\label{qpatching}
	\mathcal{T}=\begin{pmatrix}\mathds{1} & K\\0&\mathds{1}\end{pmatrix}
	\quad\mathrm{with}\quad
	K=\begin{pmatrix}0 &-N\,k &0 \\ N\,k&0&0\\0&0&0\end{pmatrix}
}
and is not contained in $G_\mathrm{geom}$.
Performing the field redefinition \eqref{fieldrefineA} we obtain
\eq{
	\widehat{G}= dx^2+dy^2+dz^2 \,,\quad
	\widehat{B}= -Nz\,dx\wedge dy \,.
}
In this frame the $Q$-flux background has a very simple form. In particular,
the metric is well-defined and the $B$-field $\widehat{B}$ just
shifts by a constant as one  moves  around the $z$-cycle. This can be compensated by
a simple gauge transformation $\widehat{B}\to
\widehat{B}+Nk\,dx\,dy$.
Besides the diffeomorphisms, the geometric group  
now contains the \td{gauge} transformations
\eqref{gengaugetrafo}. Using the $O(D,D)$-transformation \eqref{comp_66-HvM}
as well as \eqref{trantran}, the 
\td{gauge} transformations in $G_\mathrm{geom}'$ and the transition matrix \eqref{qpatching}
in the new frame read
\eq{
	\mathcal{M}_\mathsf{B}' = \begin{pmatrix} \mathds{1}&
		\widehat{G}^{-1}\mathsf{B}\widehat{G}^{-1} \\ 0 & \mathds{1}
		\end{pmatrix} \quad\mathrm{and}\quad
	\mathcal{T}'=\begin{pmatrix}  \mathds{1} &0 \\ 
		\widehat{G}K\widehat{G} & \mathds{1}
		\end{pmatrix} ,
}
respectively. Clearly,  $\mathcal{T}'=\mathcal{M}_I^{-1}\mathcal{T}\mathcal{M}_I$
is \emph{not} an element of the transformed geometric group
$G_\mathrm{geom}'$.\footnote{Note that in this example diffeomorphisms in $G_\mathrm{geom}'$
are the same as in $G_\mathrm{geom}$.} Equivalently, we observe that constant shifts 
in $\widehat{B}$
are not exact with respect to the redefined exterior derivative
\eqref{algd}. This shows that although a field redefinition is able to cast a 
non-geometric background into a simple form with transformations 
reminiscent of the usual symmetries, it
cannot provide a global description.\footnote{In \cite{Andriot:2011uh}, in addition to the
field redefinition \eqref{fieldrefineA} the further constraint
$\beta^{ij}\partial_j(\ )=0$ was  implemented which truncates the action
such that \eqref{qpatching} becomes a proper symmetry.}

To summarize, the framework we have developed can
describe non-geometric backgrounds patch-wise. If patching up requires a
transformation beyond $G_\mathrm{geom}$, different patches are still
described by different  actions \eqref{redefaction}.


\section{Conclusions}

In this paper, we have elaborated on the non-geometric part
of generalized geometry, that is, the consequences 
of the existence of $\beta$-transformations. We found  a remarkable rich
structure, which we  connected to the mathematical theory
of Lie algebroids. This  provides a general framework to study non-geometric backgrounds,
in which former studies of non-geometric actions
appear as two specific examples.

We observed that a $\beta$-transform, i.e.\
an $O(D,D)$ transformation which is not in the geometric
group, naturally gives rise to a field redefinition of
the metric and the Kalb-Ramond two-form. Expressing
the string action in these new variables, we identified the organizing principle
for the many resulting terms as the differential geometry of certain Lie algebroids.
The latter could be defined via an anchor, mapping 
either the tangent or the co-tangent space to the standard tangent bundle. 
The data of the anchor could be read off directly from the $O(D,D)$-transformation.
Particularly for $\beta$-transforms, the
Lie algebroid was not simply related to a choice of
non-holonomic basis, but gives an unprecedented
branch of differential geometry.
Note that in this latter sense,  non-geometric {\em frames} are still
geometric. 

At the heart of the paper, 
in a general setting
we proved  the connection 
between the field redefined action and the action expressed in terms of objects
appearing in the differential geometry of the associated
Lie algebroid.
Moreover, we  established  how diffeomorphisms  as well as 
gauge transformations
carry over from usual Riemannian geometry to the non-geometric
side. The behavior under
diffeomorphism originated  from the very general construction 
of the differential geometry of  the underlying Lie algebroid, 
where function-linearity was built in. 
Gauge
transformations were   more subtle as redefining with the anchor introduced a
gauge dependence in every object. To distinguish this overall gauge dependence
from an inherent gauge dependence, we  introduced the notion of a
\td{tensor}. 

We also related our non-geometric actions to double field theory.
More concretely, we showed how, for rigid $O(D,D)$ transformations,
the different non-geometric frames are related to different
solutions to the strong constraint in DFT. We confirmed
that after implementing  this solution, DFT gives indeed  
our non-geometric actions. It was fairly straightforward
to generalize the construction also to the additional
terms appearing in the effective action of superstring theory,
i.e.\ the R-R and fermionic terms. In addition, we pointed
out that higher $\alpha'$-corrections can also be described
in the non-geometric frames.

What might appear a bit disillusioning is that
these non-geometric frames do only provide a good description
of global non-geometric backgrounds in each patch.
We have seen that performing  a non-geometric field redefinition might bring 
the metric and the two-form into a very simple form.
However, the transition functions of non-geometric T-fold  
backgrounds, by definition,
involve $\beta$-transforms (i.e.\ T-duality transformations), which
are not in the symmetry group of the action in  a specific 
non-geometric frame. In other words the string action in
two patches glued together by a $\beta$-transform
cannot be described by a single non-geometric action.
Contrarily, in DFT the additional winding dependence
in the diffeomorphisms and winding diffeomorphisms
allows such a global description.


\bigskip
\noindent
\emph{Acknowledgments:}
We thank David Andriot, Gianguido Dall'Agata and Dieter L\"ust for discussions.
E.P. is supported 
by the Padova University Project CPDA105015/10
and by the MIUR-FIRB grant RBFR10QS5J.


\clearpage
\appendix


\section{Invertibility of the anchor $\gamma$}
\label{app_invertible_gamma}

In this appendix we comment on the invertibility of $\gamma$ by considering the generators of
$O(D,D)$. The Lie subgroup generated by the matrices in table
\ref{tab:transf} has dimension $2\cdot\frac{D(D-1)}{2}+D^2=2D^2-D$, which is the dimension of $O(D,D)$. Thus, this already is the identity component
$O(D,D)_0$. (Remember that a connected Lie group is generated by any open
neighborhood of the identity.) The quotient group
$O(D,D)/O(D,D)_0=\pi_0(O(D,D))=\mathbb{Z}_2\times\mathbb{Z}_2$ is generated by
the following transformations:
\eq{
  \mathcal{M}_\pm=\begin{pmatrix}
         \mathds{1}-E_1&\pm E_1\\
         \pm E_1&\mathds{1}-E_1
        \end{pmatrix},
}
where $E_1=\mathrm{diag}(1,0,\ldots,0)$. Thus, a general $O(D,D)$ matrix $\mathcal{M}$ is a (finite) product of such transformations:
\eq{
  \mathcal{M}=\left(\mathcal{M}_+\right)^{\eta_+}\left(\mathcal{M}_-\right)^{\eta_-}\prod_{i=1}^n\mathcal{M}_{\beta_i}\mathcal{M}_{\mathsf{B_i}}\mathcal{M}_{A_i},
}
where $\eta_\pm\in\{0,1\}$.
To find the frame corresponding to $\mathcal{M}$, we can apply the corresponding field redefinitions successively.

Now we just have to show that the anchor corresponding to each generator is invertible, so we consider
\eq{
	\gamma_{\mathrm{diffeo}}&=(\mathsf{A}^t)^{-1},\\
	\gamma_\mathsf{B} &= \mathds{1}, \\
	\gamma_\beta &= \mathds{1}- (G-B)\beta,\\
	\gamma_\pm &= \mathds{1}-E_1\pm(G-B)E_1.
}
For the first two, invertibility is obvious. Note that $O(D,D)$
transformations only act on the (Euclidean) compact part of the spacetime
manifold. Thus, we should be aware that the non-trivial components of the
anchor above only refer to the internal manifold, where $G$ is positive
definite. Then $(G-B)$ is positive definite (in the sense that its Hermitean
part is), and all such matrices have a positive definite inverse. So,
$\gamma_\beta=(G-B)[(G-B)^{-1}-\beta]$ is invertible as well. (Recall that
$B$ and $\beta$ are antisymmetric.) 

In order to show that $\gamma_{\pm}$ is invertible, we note that 
\eq{\det(\gamma_\pm)=\pm G_{11}=\pm\left\langle e_1, G e_1\right\rangle\neq 0,
}
where $e_1=(1,0,\ldots,0)^t$.

\clearpage
\bibliography{references}  
\bibliographystyle{utphys}


\end{document}